\documentclass[12pt]{article}

\usepackage{array,dsfont} 
\usepackage{epsfig}
\usepackage{amssymb}
\usepackage{amsmath}                          % roman hinzugefuegt
\usepackage{graphics,graphpap}

\renewcommand{\thefootnote}{\fnsymbol{footnote}}

\newcommand{\qz}{(q z)}
\newcommand{\ub}{\bar u}
\newcommand{\quark}{\langle \bar q q\rangle}
\newcommand{\mixed}{\langle \bar q \sigma gG q\rangle}
\newcommand{\squark}{\langle \bar s s\rangle}
\newcommand{\smixed}{\langle \bar s \sigma gG s\rangle}
\newcommand{\gluon}{\left\langle \frac{\alpha_s}{\pi}\,G^2\right\rangle}

\begin{document}

%%%%%%%%%% Title page
\begin{titlepage}
\begin{flushright}\begin{tabular}{l}
IPPP/05/76\\
DCPT/05/152
\end{tabular}
\end{flushright}
\vskip1.5cm
\begin{center}
   {\Large \bf \boldmath Operator Relations for SU(3) Breaking\\[5pt]
     Contributions to K and K$^*$  Distribution Amplitudes}
    \vskip1.3cm {\sc
Patricia Ball\footnote{Patricia.Ball@durham.ac.uk} and Roman 
Zwicky\footnote{Roman.Zwicky@durham.ac.uk}
  \vskip0.5cm
        {\em IPPP, Department of Physics, 
University of Durham, Durham DH1 3LE, UK}} \\
\vskip2.5cm 

%{\em Version of \today}

\vskip3cm

{\large\bf Abstract:\\[10pt]} \parbox[t]{\textwidth}{
We derive constraints on the asymmetry $a_1$ of the momentum fractions
carried by quark and antiquark in $K$
and $K^*$ mesons in leading twist. These constraints follow from
exact operator identities and relate $a_1$ to SU(3) breaking
quark-antiquark-gluon matrix elements which we determine from QCD sum
rules. Comparing our results to determinations of $a_1$ from QCD sum
rules based on correlation functions of quark currents, we find that,
for $a_1^\parallel(K^*)$ the central values agree well and come with
moderate errors, whereas for $a_1(K)$ and $a_1^\perp(K^*)$ the results
from operator relations are consistent with those from quark current
sum rules, but come with larger uncertainties. The consistency of
results confirms that the QCD sum rule method is indeed suitable for the
calculation of $a_1$. We conclude that 
the presently most accurate predictions for $a_1$ come from the direct
determination from QCD sum rules based on correlation functions of
quark currents and are given by:
$$
a_1(K) = 0.06\pm 0.03, \quad a_1^\parallel(K^*) = 0.03\pm 0.02,\quad 
a_1^\perp(K^*) = 0.04\pm 0.03.
$$
}

\vfill

%{\em submitted to Physics Letters B}
\end{center}
\end{titlepage}

\setcounter{footnote}{0}
\renewcommand{\thefootnote}{\arabic{footnote}}

\newpage

\section{Introduction}\label{sec:1}

Hadronic light-cone distribution amplitudes (DAs) of 
leading twist have been attracting considerable interest in the
context of B physics. They enter the amplitudes of QCD processes
that can be described in collinear factorisation, which include, 
to leading order in an expansion in $1/m_b$, a large class
of nonleptonic B decays \cite{BBNS}, such as $B\to\pi\pi,KK$. 
DAs are also an essential ingredient in the calculation of weak decay
form factors such as $B\to\pi,\rho,K,K^*$ from QCD sum rules on the
light-cone \cite{FFs}.
These decays, and their CP asymmetries, are currently being studied at 
the B factories
BaBar and Belle and are expected to yield essential information about the
pattern of CP violation and potential sources of flavour violation
beyond the SM. 

One particular problem in this context is the size of SU(3) breaking
corrections to $K$ and $K^*$ DAs, which has been studied in a number
of recent papers \cite{elena,alex,lenz,BZ}. 
%The most relevant SU(3)
%breaking quantity is $a_1 = \frac{5}{3}\,\langle u_s-u_q\rangle$, that
%is the average asymmetry of the momentum fractions $u_s$ and $u_q$
%carried by, respectively, the $s$ quark and the light antiquark $\bar
%q$ in the meson. All existing calculations rely on QCD sum rules
The DAs themselves are  defined as matrix elements of
  quark-antiquark gauge-invariant nonlocal operators on the light-cone.
To leading-twist accuracy, there are three such DAs for $K$ and $K^*$
  ($z^2=0$):
  \begin{eqnarray}
    \langle 0 |\bar q(z)\slash\!\!\! z \gamma_5 [z,-z]s(-z)
  |K(q)\rangle &=& i  f_K \qz 
  \int_0^1 du\, e^{i\xi\qz} \phi_K(u)\,,\nonumber\\
  \langle 0 |\bar q(z)\slash\!\!\! z [z,-z] s(-z)
  |K^{*}(q,\lambda)\rangle &=& (e^{(\lambda)} z)
  f_K^\parallel m_{K^*}\int_0^1 du\, e^{i\xi\qz}
  \phi_K^\parallel(u),\nonumber\\
  \langle 0 |\bar q(z)\sigma_{\mu\nu}[z,-z]s(-z)
  |K^{*}(q,\lambda)\rangle & = &
  i(e^{(\lambda)}_\mu q_\nu -e^{(\lambda)}_\nu q_\mu)
  f_K^\perp(\mu) \int_0^1 du\, e^{i\xi\qz} \phi_K^\perp(u),
  \label{eq:defDAs}
  \end{eqnarray}
  with the Wilson-line
  $$
  [z,-z] = \mbox{Pexp}\,\left[ig\int_{-1}^1 d\alpha\, z^\mu A_\mu(\alpha
    z)\right]
  $$
inserted between quark fields to render the matrix elements
gauge-invariant. 
  In the above definitions, $e^{(\lambda)}_\nu$ is the
   polarisation vector of a vector meson with polarisation
   $\lambda$; there are two leading-twist 
DAs for vector mesons, $\phi_K^\parallel$
   and $\phi_K^\perp$, corresponding to
   longitudinal and transverse polarisation, respectively.
 The integration variable $u$ is the (longitudinal)
meson momentum fraction carried by the quark, $\ub \equiv 1-u$ the
  momentum fraction carried by the antiquark and $\xi = u-\ub$.  
The decay constants
  $f_K^{(\parallel,\perp)}$ are defined in the usual way by the local limit of
  Eqs.~(\ref{eq:defDAs}) and chosen in such a way that 
  \begin{equation}\label{eq:norm} \int_0^1 du\, \phi(u)=1.\end{equation}
   All three distributions $\phi_K,
  \phi_K^\parallel , \phi_K^\perp$ can be expanded in Gegenbauer
  polynomials $C_n^{3/2}$,
\begin{equation}
\phi(u) = 6u\bar u\left(1 + \sum_{n \geq 1} a_n C_n^{3/2}(2u-1)\right)\,,
\end{equation}
where the $a_n$ are hadronic parameters, the so-called Gegenbauer moments.

The most relevant quantities characterising SU(3) breaking of these
DAs are the
decay constants $f_K$ and $f_{K}^{\perp,\parallel}$, and $a_1(K)$ and
$a_1^{\perp,\parallel}(K^*)$, which can be expressed in terms of the
DAs as
\begin{equation}
a_1(K) = \frac{5}{3}\, \int_0^1 du\,
 (u-\bar u)\phi_K(u)
\end{equation}
and correspondingly for $a_1^{\parallel,\perp}(K^*)$.
$a_1$ describes the
difference of the average longitudinal momenta of the quark and
antiquark in the two-particle Fock-state component of the meson, a quantity
that vanishes for particles with equal-mass quarks (particles with 
definite G-parity). The decay constants $f_K$ and
$f_{K}^\parallel$ can be extracted from experiment;
$f_{K}^\perp$ has been calculated  from both lattice \cite{lattbec}
and QCD sum rules, e.g.\ Ref.~\cite{BZ}. 
In this paper we focus on the determination of $a_1$: no lattice
calculation of this quantity has been attempted yet, so essentially
all available information on $a_1$ comes from QCD sum rule
calculations. $a_1$ can be calculated either directly from the
correlation function of two quark currents 
\cite{elena,alex,BZ,Russians,CZreport} or from  operator
identities relating it to certain quark-quark-gluon matrix elements,
denoted $\kappa_4$, which  
are calculated from QCD sum rules themselves \cite{lenz}. In a
previous paper, Ref.~\cite{BZ}, we have obtained the following results
from the first method, at the scale of 1~GeV:
\begin{equation}
a_1(K)^{\rm BZ} = 0.050\pm 0.025,\quad a_1^\parallel(K^*)^{\rm BZ} = 0.025\pm
0.015,\quad a_1^\perp(K^*)^{\rm BZ} = 0.04\pm 0.03,
\end{equation}
whereas Braun and Lenz found the following results from
operator identities  \cite{lenz}:
\begin{equation}\label{reslenz}
a_1(K)^{\rm BL} = 0.10\pm 0.12,\qquad a_1^\parallel(K^*)^{\rm BL} = 
0.10\pm 0.07.
\end{equation}
These results were obtained to first order in $m_s$ and neglecting
explicit terms in $m_s^2$ and $m_q$ in the operator identities. 
Numerically, however, these terms are not negligible: the
$O(m_s^2)$ correction shifts $a_1(K)$ by $+0.17$ and
  $a_1^\parallel(K^*)$ by $+0.08$ for our central value of
  $m_s$. Corrections in $m_q$ are small for $a_1^\parallel(K^*)$, but
  chirally enhanced for $a_1(K)$ and shift $a_1(K)$ by $+0.04$ for our
  central value of $m_q$. A consistent inclusion of $O(m_{q,s})$ effects
  requires the calculation of these terms also for $\kappa_4$. In the
  present paper, we present such a calculation and  improve the sum
  rules for $\kappa_4$ derived in Ref.~\cite{lenz}
by the inclusion of all dominant terms to $O(m_q^2)$ and $O(m_s^2)$,
which include in particular
two-loop perturbative and gluon-condensate
contributions. The 
perturbative contributions come with large
coefficients and prove to be very relevant numerically. We then
construct several sum rules for $\kappa_4$ which differ
by the chirality structure of the involved currents and the
spin-parity assignment of the hadronic states coupling to them.
We provide criteria that allow one to identify the sum rules most
suitable for the calculation of $\kappa_4$ and obtain the
corresponding numerical results, including a careful
analysis of the theoretical uncertainty of $\kappa_4$ and the
corresponding values of $a_1$. One important finding of our paper is
that the results of these calculations agree, within errors, with
those from the quark current sum rules, which shows 
that the application of the QCD sum rule method to the calcualation of
$a_1$ yields mutual consistent results. It is this consistency that
strengthens our confidence in the validity of the results for $a_1$.

Our paper is organised as follows: in Sec.~\ref{sec:2} we derive
the operator relations for $a_1$, in Sec.~\ref{sec:3} we obtain numerical
results for the corresponding matrix elements and compare with the
results  of Ref.~\cite{BZ}. In Sec.~\ref{sec:4}
we summarise and conclude. The paper also contains two appendices
giving explicit expressions 
for all relevant correlation functions and Borel transforms.

\section{\boldmath Exact Identities for 
$a_1$}\label{sec:2}

In Ref.~\cite{lenz}, the following relations were obtained for $a_1(K)$
and $a_1^\parallel(K^*)$:
\begin{eqnarray}
\frac{9}{5}\, a_1(K) & = & -\frac{m_s-m_q}{m_s+m_q} +
4\,\frac{m_s^2-m_q^2}{m_{K}^2}  - 8 \kappa_{4}(K)\,,\label{a1K}\\
\frac{3}{5}\,a_1^\parallel(K^*) &=&
-\frac{f_K^\perp}{f_K^\parallel}\,\frac{m_s-m_q}{m_{K^*}} + 2 
\,\frac{m_s^2-m_q^2}{m_{K^*}^2} - 4 \kappa_{4}^\parallel(K^*),\label{a1Kpar}
\end{eqnarray}
where $\kappa_4(K)$ and $\kappa_4^\parallel(K^*)$ are twist-4
quark-quark-gluon  matrix elements defined by
\begin{eqnarray}
\label{k4}
\langle 0 | \bar q  (g G_{\alpha\mu}) 
i\gamma^\mu\gamma_5  s|K(q)\rangle &=& i q_\alpha f_{K} m_K^2 \kappa_{4}(K),\\
\label{k4L}
\langle 0 | \bar q (g G_{\alpha\mu}) i\gamma^\mu   s | K^*(q,\lambda) \rangle
&=& e^{(\lambda)}_\alpha  f_K^\parallel m_{K^*}^3 \kappa^\parallel_{4}(K^*).
\end{eqnarray}
{$\kappa_4(K)$ and $\kappa_4^\parallel(K^*)$
vanish for  $m_s\to m_q$ due to G-parity. 
The special structure of (\ref{a1K}) allows one to
determine the value of $\kappa_4(K)$ to leading order in $m_s$ for
$m_q\to 0$ \cite{lenz},
\begin{equation}\label{10}
\kappa_4(K) = -\frac{1}{8},
\end{equation}
which is a consequence of the conservation of the axial current in the 
chiral limit.

The above relations were derived from the analysis of matrix elements
of the local operators 
($\stackrel{\leftrightarrow}{D} = \stackrel{\rightarrow}{D} -
\stackrel{\leftarrow}{D}$)
\begin{equation}
O_{\mu\nu}^{(5)} = \frac{1}{2}\,\bar q \gamma_\mu(\gamma_5) i
\stackrel{\leftrightarrow}{D}_\nu s + \frac{1}{2}\,\bar q 
\gamma_\nu(\gamma_5) i
\stackrel{\leftrightarrow}{D}_\mu s - 
\frac{1}{4}\,g_{\mu\nu} \bar q i(\gamma_5)
\stackrel{\leftrightarrow}{\slash\!\!\!\!  D} s,
\end{equation}
whose divergence can be expressed in terms of bilinear quark
operators. In this section, we rederive these relations in a different
way and obtain a new relation for $a_1^\perp(K^*)$.

The starting point for our analysis are the exact nonlocal operator
relations \cite{BB98,PB98}
\begin{eqnarray}
\frac{\partial}{\partial x_\mu}\, \bar q(x)\gamma_\mu (\gamma_5) s(-x)
& = &{} - i \int_{-1}^1 dv\, v \bar q (x) x_\alpha
gG^{\alpha\mu}(vx) \gamma_\mu(\gamma_5) s(-x)\nonumber\\
&&{} - (m_s\pm m_q) \bar
q(x)i(\gamma_5) s(-x),\label{I}\\
\partial^\mu \{\bar q(x)\gamma_\mu(\gamma_5) s(-x)\}
& = & {}\! - i\!\int_{-1}^1 dv\, \bar q(x) x_\alpha
G^{\alpha\mu}(vx) \gamma_\mu(\gamma_5) s(-x)\nonumber\\
&&{}- (m_q\mp m_s) \bar
q(x)i(\gamma_5) s(-x),\label{II}
\end{eqnarray}
where the total translation $\partial_\mu$ is defined as
\begin{equation}
\partial_\mu \left\{ \bar q(x)\Gamma s(-x)\right\} \equiv
\left.\frac{\partial}{\partial y_\mu}\,\left\{ \bar q(x+y) [x+y,-x+y]
    \Gamma s(-x+y)\right\}\right|_{y\to 0}.
\end{equation}
The corresponding nonlocal  matrix elements are, for $K$ and
$K^*_\parallel$ ($x^2\neq 0$):
\begin{eqnarray}
\langle 0 | \bar q(x)\gamma_\mu\gamma_5 s(-x)|K(q)\rangle\
& = & i f_K q_\mu \int_0^1 du \, e^{i\xi qx} \left[ \phi_K(u) +
  O(x^2)\right]\nonumber\\
&&{} + \frac{i}{2}\,f_K m_K^2\,
  \frac{1}{qx}\, x_\mu \int_0^1 du \, e^{i\xi qx}\,
\left[g_K(u)-\phi_K(u) + O(x^2)\right]\!,\label{12}\\
\langle 0 | \bar q(x)i\gamma_5 s(-x)|K(q)\rangle\
& = & \frac{f_K m_K^2}{m_s+m_q}\,\int_0^1 du \, e^{i\xi
  qx}\,\left(\phi_K^p(u)+O(x^2)\right),\label{13}\\
\langle 0|\bar q(x) \gamma_\mu s(-x)|K^*(q,\lambda)\rangle
 &=& f_K^\parallel m_{K^*} \Bigg\{
\frac{e^{(\lambda)}x}{qx}\, q_\mu \int_0^1 du \,e^{i\xi qx}
\Big[\phi_K^\parallel(u)+O(x^2)\Big]\nonumber\\
&&{}+\left(e^{(\lambda)}_\mu-q_\mu\frac{e^{(\lambda)}x}{qx}\right)
\int_0^1 du\, e^{i\xi qx} \left(g_K^v(u)+O(x^2)\right)
\nonumber\\&&{}\hspace*{-2cm}
-\frac{1}{2}x_\mu \frac{e^{(\lambda)}x}{(qx)^2} m^2_{K^*}\!\! \int_0^1
\! du
\, e^{i\xi qx} \left[g_{K}^{(3)}(u)+\phi_K^\parallel(u)-2 g_K^v(u)+
O(x^2)\right]\!
\!\Bigg\}.\hspace*{15pt}\label{14}
\end{eqnarray}
In the above definitions, $\phi_K$ and
$\phi_K^\parallel$ are the leading-twist DAs of $K$ and
$K_\parallel^*$, respectively; all other functions are
higher-twist DAs and have been extensively discussed in
Refs.~\cite{BB98,PB98,BBKT,prep}. 

$a_1(K)$, the quantity we are interested in, is related to the first
moment of $\phi_K$:  
$$a_1(K)=\frac{5}{3}M_1^{\phi_K}$$ 
with $M_1^f \equiv \int_0^1 du\, (u-\bar u) f(u)$
being the first moment of the DA $f(u)$. 
Taking the matrix elements of (\ref{I}) and (\ref{II}) for $K$ and 
expanding to leading order in $x^2$ and next-to-leading order in $qx$, 
one obtains the exact relations 
\begin{eqnarray}
M_1^{\phi_K} - 2 M_1^{g_K} & = &
-\frac{m_s-m_q}{m_s+m_q},\nonumber\\
\frac{1}{2}\,\left( M_1^{\phi_K} + M_1^{g_K}\right)& = &
-2 \kappa_4(K) + M_1^{\phi_K^p},\label{sysK}
\end{eqnarray}
from which one can determine $M_1^{\phi_K}$ once either
$M_1^{\phi_K^p}$ or $M_1^{g_K}$ are known. $g_K$ is a twist-4 DA and
$ M_1^{g_K}$ contains quark-quark-gluon matrix elements itself, cf.\
Refs.~\cite{PB98,prep}, whereas $\phi_K^p$ is twist-3 and
$M_1^{\phi_K^p}$ is
completely determined in terms of the twist-2 DA $\phi_K$ and mass
corrections. $M_1^{\phi_K^p}$ can be obtained from a second set of
nonlocal operator relations involving tensor currents $\bar q(x)\sigma_{\mu\nu}
\gamma_5s(-x)$ or, equivalently, 
from the recursion relations for the moments of $\phi_K^p$
given in Ref.~\cite{prep}: 
$$M_1^{\phi_K^p} = \frac{m_s^2-m_q^2}{m_K^2}.$$
Solving (\ref{sysK}) for $a_1(K)$, we then rederive 
\begin{equation}
\frac{9}{5}\, a_1(K)  =  -\frac{m_s-m_q}{m_s+m_q} +
4\,\frac{m_s^2-m_q^2}{m_{K}^2}  - 8 \kappa_{4}(K)\,,
\end{equation}
which confirms the result obtained in Ref.~\cite{lenz}. Note that the
first term on the right-hand side is rather sensitive to the value
of $m_q$ and the second one to that of $m_s$. 

For $K^*_\parallel$, the same method yields the equations
\begin{eqnarray}
 M_1^{\phi_K^\parallel} + M_1^{g_K^{(3)}} & = &2 M_1^{g_K^v},\nonumber\\
 M_1^{\phi_K^\parallel} - M_1^{g_K^{(3)}} & = & -2\,
 \frac{f_K^\perp}{f_K^\parallel}\,\frac{m_s-m_q}{m_{K^*}} + 2\,
 \frac{m_s^2-m_q^2}{m_{K^*}^2} - 4 \kappa_4^\parallel(K^*).\label{syspar}
\end{eqnarray}
Again, $g_K^{(3)}$ is a twist-4 DA whose first moment is not known from
  any independent analysis, whereas 
$M_1^{g_K^v}$, the first moment of the twist-3 DA $g_K^v$, 
can be read off Eq.~(4.6) in Ref.~\cite{BBKT}:
\begin{equation}\label{gv}
2 M_1^{g_K^v} = M_1^{\phi_K^\parallel} +
\frac{f_K^\perp}{f_K^\parallel}\, \frac{m_s-m_q}{m_{K^*}}.
\end{equation}
We can then solve (\ref{syspar}) for 
$a_1^\parallel(K^*)$ and obtain
\begin{equation}\label{rela1par}
\frac{3}{5}\,a_1^\parallel(K^*) =
-\frac{f_K^\perp}{f_K^\parallel}\,\frac{m_s-m_q}{m_{K^*}} + 2 
\,\frac{m_s^2-m_q^2}{m_{K^*}^2} - 4 \kappa_4^\parallel(K^*),
\end{equation}
which agrees with Eq.~(\ref{a1Kpar}), the result obtained in  Ref.~\cite{lenz}.

Let us now apply the same method to chiral-odd operators, with the aim
of obtaining an analogous new expression for $a_1^\perp(K^*)$. The
relevant nonlocal operator relations are
\begin{eqnarray}
\lefteqn{\frac{\partial}{\partial x_\mu} \bar q(x) \sigma_{\mu\nu} s(-x)
  =  -i \partial_\nu \bar q(x) s(-x)
+ (m_s-m_q)  \bar q(x) \gamma_\nu s(-x)}\hspace*{2cm}\nonumber\\
&&{}+\int_{-1}^1 dv \bar q(x) g x_\alpha
 G^{\alpha}_{\phantom{\alpha}\nu}(vx) s(-x) - i 
\int_{-1}^1 dv v \bar q(x) g x_\alpha G^{\alpha\mu}(vx)\sigma_{\mu\nu}
s(-x),\nonumber
\end{eqnarray}
\begin{eqnarray}
\partial^\mu \{ \bar q(x) \sigma_{\mu\nu} s(-x) \} & = &
-i\frac{\partial}{\partial x_\nu}\, \bar q(x) s(-x) - 
(m_s+m_q) \bar q(x) \gamma_\nu s(-x)\nonumber\\
&&\hspace*{-10pt}{}+\int_{-1}^1 dv v \bar q(x) g x_\alpha
 G^{\alpha}_{\phantom{\alpha}\nu}(vx) s(-x) - i 
\int_{-1}^1 dv \bar q(x) g x_\alpha G^{\alpha\mu}(vx)\sigma_{\mu\nu}
s(-x).\nonumber\\[-20pt]\label{eq:OPrel}
\end{eqnarray}
These relations were first derived, without the terms in $m_s\pm
m_q$, in Ref.~\cite{BB98}; the terms in the quark masses are new.

The relevant $K^*$ matrix elements are given by \cite{BB98}:
$$
\langle 0|\bar q(x) \sigma_{\mu \nu}
s(-x)|K^*(q,\lambda)\rangle = 
 i f_K^\perp \left[ (e^{(\lambda)}_{\mu}q_\nu -
e^{(\lambda)}_{\nu}q_\mu )
\int_{0}^{1} \!du\, e^{i \xi q x}
\Bigg[\phi_K^{\perp}(u) + O(x^2)\Bigg] \right.
$$
\begin{eqnarray}
& &{}+ (q_\mu x_\nu - q_\nu x_\mu )
\frac{e^{(\lambda)} x}{(q x)^{2}}m_{K^*}^{2}
\int_{0}^{1} \!du\, e^{i \xi q x} \left[ h_K^t(u) -
\frac{1}{2}\,\phi_K^\perp(u) - \frac{1}{2}\,h_K^{(3)}(u) + O(x^2)\right]
\nonumber \\
& & \left.{}+ \frac{1}{2}
(e^{(\lambda)}_{ \mu} x_\nu -e^{(\lambda)}_{ \nu} x_\mu)
\frac{m_{K^*}^{2}}{q  x}
\int_{0}^{1} \!du\, e^{i \xi q x} \left( h_K^{(3)}(u)-
\phi_K^\perp(u) + O(x^2)\right)  \right],
\label{eq:OPE2}
\end{eqnarray}
\begin{eqnarray}
\lefteqn{\langle 0 | \bar q(x) s(-x) | K^*(q,\lambda)\rangle
  =}\hspace*{2cm}\nonumber\\&&{}= -i
\left(f_K^\perp - f_K^\parallel\,\frac{m_s+m_q}{m_{K^*}}\right)
\left(e^{(\lambda)} x\right) m_{K^*}^2 \int_0^1 du\, e^{i\xi qx}
\left(h_K^s(u) + O(x^2)\right),
\end{eqnarray}
where, again, $\phi_K^\perp$ is the leading-twist DA of the
transversely polarised $K^*$ and $h_K^{s,t}$
and $h_K^{(3)}$ are higher-twist DAs. 
In addition, we also need the following quark-quark-gluon matrix element:
\begin{eqnarray}
\langle 0 | \bar q (g G_{\alpha}^{\phantom{\alpha}\mu})
\lefteqn{\sigma_{\beta\mu}s|K^*(q,\lambda)\rangle =}\nonumber\\
& = & f_K^\perp m_{K^*}^2 \!\left\{\frac{1}{2}\,\kappa_3^\perp(K^*)
(e^{(\lambda)}_\alpha q_\beta + e^{(\lambda)}_\beta q_\alpha) + 
\kappa_4^\perp(K^*)
(e^{(\lambda)}_\alpha q_\beta - e^{(\lambda)}_\beta q_\alpha)\right\}.
\label{def:T1T2}
\end{eqnarray}
Here $\kappa_3^\perp(K^*)$ is a twist-3 matrix element,
$\kappa_4^\perp(K^*)$ is twist-4; both are $O(m_s-m_q)$ due to
G-parity.\footnote{The normalisation of $\kappa_3^\perp(K^*)$ is chosen 
in such a way that
  $\int{\cal D}\underline{\alpha} {\cal T}(\underline{\alpha})=
  \kappa_3^\perp(K^*)$ for the twist-3 DA ${\cal
    T}(\underline{\alpha})$ defined in Ref.~\cite{BBKT}.}
Taking matrix elements of (\ref{eq:OPrel}), one obtains expressions in
$q_\nu$, $e^{(\lambda)}_\nu$ and $x_\nu$. To twist-4 accuracy only the
former two are relevant and yield a set of four linear 
equations for the four first moments of $g_K^v$,
$h_K^s$, $h_K^t$ and $h_K^{(3)}$:
\begin{eqnarray}
-(\kappa_3^\perp(K^*)-2\kappa_4^\perp(K^*)) 
+ \delta_+ M_1^{g_K^v} + M_1^{h_K^s} & = & \frac{1}{2}\,M_1^{h_K^{(3)}} +
 \frac{1}{2} \, M_1^{\phi_K^\perp}\,,\nonumber\\
\kappa_3^\perp(K^*)+ 2\kappa_4^\perp(K^*) + 
\delta_+ M_1^{g_K^v} - M_1^{h_K^s} - \delta_+
 M_1^{\phi_K^\parallel} & = &  
\frac{1}{2}\, M_1^{h_K^{(3)}} - M_1^{h_K^t} + \frac{1}{2}\,
M_1^{\phi_K^\perp}\,,\nonumber\\
3\, M_1^{h_K^{(3)}} - \,M_1^{\phi_K^\perp} & = &
2 \delta_-\,,\nonumber\\
M_1^{h_K^3} - 2 M_1^{h_K^t} + M_1^{\phi_K^\perp} & = & 0
\end{eqnarray}
with $\delta_\pm = \frac{f_K^\parallel}{f_K^\perp}\,\frac{m_s\pm m_q}{m_K^*}$.
The solution of that system implies
$$
\delta_+ M_1^{g_K^v} = \frac{1}{6}\,\delta_- + \frac{1}{2}\, 
\delta_+ M_1^{\phi_K^\parallel}
+ \frac{1}{3}\, M_1^{\phi_K^\perp} - 2\kappa_4^\perp(K^*),
$$
which must agree with $M_1^{g_K^v}$ as given in Eq.~(\ref{gv}). 
Solving for $a_1^{\perp}(K^*)$, one finds
\begin{equation}\label{eq:perp}
\frac{3}{5}\, a_1^\perp(K^*) =
-\frac{f_K^\parallel}{f_K^\perp}\,\frac{m_s-m_q}{2m_{K^*}} +
  \frac{3}{2}\,\frac{m_s^2-m_q^2}{m_{K^*}^2} + 6 \kappa_4^\perp(K^*),
\end{equation}
which is the wanted new relation for $a_1^\perp(K^*)$. Note that in all
three relations (\ref{a1K}), (\ref{a1Kpar}) and (\ref{eq:perp})
$\kappa_4$ enters multiplied with a large numerical factor which
implies that the theoretical uncertainty of the resulting values of
$a_1$ will be much larger than that of $\kappa_4$ itself.

\section{\boldmath QCD Sum Rules for $\kappa_4$, $\kappa_4^\parallel$
  and $\kappa_4^\perp$}\label{sec:3}

In order to obtain numerical predictions for $a_1$ from the
relations derived in the last section, one needs to know the values of
the $\kappa_4$ matrix elements. 
$\kappa_4(K)$ and $\kappa_4^\parallel(K^*)$ have been calculated in
Ref.~\cite{lenz} from QCD sum rules to leading order in SU(3)
breaking parameters with the following results:
\begin{equation}\label{28}
\kappa_4(K)^{\rm BL}= -0.11\pm 0.03,\qquad \kappa_4^\parallel(K^*)^{\rm BL} =
-0.050\pm 0.010,
\end{equation}
which, using the relations (\ref{a1K}) and (\ref{a1Kpar}), letting
$m_q=0$ and neglecting the terms in $m_s^2$ 
translates into \cite{lenz}
\begin{equation}\label{29}
a_1(K)^{\rm BL} = 0.10 \pm 0.12, \qquad a_1^\parallel(K^*)^{\rm BL} = 
0.10\pm 0.07.
\end{equation}
All these results refer to a renormalisation scale of 1~GeV. 

\begin{table}[bt]
\renewcommand{\arraystretch}{1.3}
\addtolength{\arraycolsep}{3pt}
$$
\begin{array}{|r@{\:=\:}l||r@{\:=\:}l|}
\hline 
\quark & (-0.24\pm0.01)^3\,\mbox{GeV}^3 & \squark & (1-\delta_3)\,\quark\\
\mixed & m_0^2\,\quark &  \smixed & (1-\delta_5)\mixed\\[6pt]
\displaystyle \gluon & (0.012\pm 0.003)\, 
{\rm GeV}^4 & \multicolumn{2}{l|}{}\\[6pt]\hline
\multicolumn{4}{|c|}{m_0^2 = (0.8\pm 0.1)\,{\rm GeV}^2,\quad \delta_3
  = 0.2\pm 0.2, \quad \delta_5 = 0.2\pm 0.2}\\\hline
\multicolumn{4}{|c|}{\overline{m}_s(2\,\mbox{GeV}) = (100\pm
20)\,\mbox{MeV}~~~\longleftrightarrow~~~ \overline{m}_s(1\,\mbox{GeV})
= (137\pm 27)\,\mbox{MeV}}\\
\multicolumn{4}{|c|}{\overline{m}_q(\mu) = \overline{m}_s(\mu)/R,
  \quad R = 24.4\pm 1.5}\\\hline
\multicolumn{4}{|c|}{\alpha_s(m_Z) = 0.1187\pm 0.002  ~\longleftrightarrow~ 
\alpha_s(1\,\mbox{GeV}) = 0.534^{+0.064}_{-0.052}}\\\hline
\multicolumn{4}{|c|}{f_K = (0.160\pm 0.002)\,{\rm GeV},\qquad f_K^\parallel =
  (0.217\pm 0.005)\,{\rm GeV}}\\
\multicolumn{4}{|c|}{ f_K^\perp = (0.185\pm0.010)\,{\rm
    GeV}}\\\hline
\end{array}
$$
\renewcommand{\arraystretch}{1}
\addtolength{\arraycolsep}{-3pt}
\vskip-10pt
\caption[]{Input parameters for sum rules at the
  renormalisation scale $\mu=1\,$GeV. The value of $m_s$ is obtained
  from 
  unquenched lattice calculations with $n_f=2$ flavours 
as summarised in \cite{mslatt}, which agrees with the results from QCD
  sum rule calculations \cite{jamin}. $\overline{m}_q$ is taken from
  chiral perturbation theory \cite{chPT}.\footnotemark[2] 
$\alpha_s(m_Z)$ is the PDG
  average \cite{PDG}, $f_K$ and $f_K^\parallel$ are
  known from experiment and $f_K^\perp$ has been determined in
  Refs.~\cite{BZ,lattbec}. The errors of quark masses and condensates
  are treated as correlated, see text.}\label{tab:1}
\end{table}

\footnotetext[2]{ $m_q$ has also been
  determined from lattice calculations. The most recent papers on this
  topic are Refs.~\cite{mqlatt}. The central value of $m_s/m_q$
  determined in the first of these papers with $n_f=2$ running
  flavours and nonperturbative renormalisation agrees with the result
  from chiral perturbation theory, whereas the result of the second, 
obtained with $n_f=3$  and perturbative (two-loop) renormalisation, 
is a bit lower. As the field appears to develop rapidly, we
  refrain from taking either side and stay with the
  result from chiral perturbation theory.}\addtocounter{footnote}{1}

In this section we present QCD sum rules for $\kappa_4(K)$ and
$\kappa_4^\parallel(K^*)$ which  
are accurate to NLO in SU(3) breaking and also a new sum rule for
$\kappa_4^\perp(K^*)$ to the same accuracy. For all sum rules we
include $O(m_q)$ effects.
The sum rules are of the generic form
\begin{equation}\label{generic}
\kappa_4(K) f_K^2 m_K^n e^{-m_K^2/M^2} + \mbox{contribution from 
higher mass states} = {\cal B}_{M^2}\Pi_G,
\end{equation}
and correspondingly for $K^*$. $\Pi_G$ are correlation functions of type
$$
\Pi_G(q) = i\int d^4y e^{iqy} \langle 0 | T [\bar q (g
  G_{\alpha\mu})\Gamma_1^\mu  s](y) 
[\bar s \Gamma_2 q](0)|0\rangle
$$
with suitably chosen Dirac structures $\Gamma_1^\mu$ and $\Gamma_2$;
explicit expressions for all relevant $\Pi_G$ are given in App.~\ref{appA}.
${\cal B}_{M^2}\Pi_G$ is the Borel transform of $\Pi_G$, $M^2$ the Borel
  parameter and $n$ is either 2 or 4. In order to separate the ground
  state from higher mass contributions, one usually models the latter,
  using global quark hadron duality,
  by an integral over the perturbative spectral density:
\begin{equation}\label{contmodel}
\mbox{contribution from 
higher mass states} \approx \int_{s_0}^\infty e^{-s/M^2}
\,\frac{1}{\pi} \,{\rm Im}\,\Pi_G(s);
\end{equation}
the parameter $s_0$ is called continuum threshold. The input
parameters for the QCD sum rules are collected in Tab.~\ref{tab:1}. 
%As $\kappa_4$ are SU(3) breaking parameters, we expect them to be 
%sensitive to SU(3) breaking in QCD sum rules, in particular the
%parameters $\delta_3$ and $\delta_5$,
%which come with a rather large uncertainty. In addition, as
%$\kappa_4$ are quark-quark-gluon matrix elements, we also expect them
%to be sensitive to $\alpha_s$ at the hadronic scale $\sim\,$1~GeV. This
%implies that our final results for $\kappa_4$ will come with
%nonnegligible uncertainty and induce even larger uncertainties for
%$a_1$. 

All $\kappa_4$ parameters can actually be determined from more than one sum
rule derived from various $\Pi_G$ which
can be characterised by the following features:
\begin{itemize}
\item the currents can have the same or different chirality, which
  results in chiral-even and chiral-odd sum rules, respectively;
\item the hadronic states saturating $\Pi_G$ can have unique spin-parity
 or come with different parity (e.g.\ $0^-$ and $1^+$),
 which results in pure-parity and mixed-parity sum rules, respectively.
\end{itemize}
Note that all chiral-odd sum rules are also pure-parity.

In chiral-odd sum rules the quark condensates always appear in the
combination $\quark-\squark = \delta_3\quark$ and $\mixed-\smixed =
\delta_5 \mixed$, which induces a large dependence on the only
poorly constrained parameters $\delta_{3,5}$ 
and also increases the impact of the
gluon condensate contribution which is equally poorly known. We
therefore decide to drop all chiral-odd sum rules and only use
chiral-even ones.

As for mixed and pure-parity sum rules, they come with different mass
dimensions:  $n=2$ in (\ref{generic}) for mixed-parity 
vs.\ $n=4$ for pure-parity sum rules. 
It is an important result of this paper that the continuum
contributions to the mixed-parity sum rules,
for typical Borel parameters $M^2$ around $1.7\, {\rm GeV^2}$, are
small and below 10\% for all three $\kappa_4$. Pure-parity sum
rules, on the other hand, have a large continuum contribution around 
30\%. There are two reasons for this result: first, the additional power
of $m_K^2$ in pure-parity sum rules counteracts the exponential suppression
of the continuum contribution. Second, the contributions of particles
with different parity have different sign: it was already found in 
Ref.~\cite{lenz} 
that $\kappa_4(K)$ and $\kappa_4^\parallel(K_1)$ have 
opposite sign; we find that the same applies to
$\kappa_4^\parallel(K^*)$ and the
corresponding  $\kappa_4(K^*_0)$ of the lowest scalar resonance, and
ditto to $\kappa_4^\perp(K^*)$ and the coupling
$\kappa_4^\perp(K_1)$ of the axial vector $K_1$ meson. 
These results suggest that the $\kappa_4$
matrix elements of opposite-parity mesons have
generically different signs and tend to cancel each other in mixed-parity sum
rules, which results in a small continuum contribution. From a more
formal point of view it is rather obvious from the definitions
Eqs.~(\ref{k4}), (\ref{k4L}) and (\ref{def:T1T2}) that the sign of
$\kappa_4$ changes under a parity transformation,\footnote{In QCD parity
is not a symmetry of the hadronic spectrum because the 
$U(1)_{\rm A}$-symmetry is broken.}
which is in line with our findings.

The mixed-parity sum rules for $K$ and $K^*$ do involve the three 
spin-parity systems $(0^-,1^+)$, $(1^-,0^+)$ and $(1^-,1^+)$. Note
that for all of them  the ``wrong''-parity ground state (e.g.\ the scalar 
$K^*_0(1430)$) and the first
orbital excitation of the ``right''-parity state (e.g.\ the vector
$K^*(1410)$) have nearly equal mass, which makes the cancellation very
effective. 
We conclude that  mixed-parity sum rules are more
 reliable than pure-parity ones and, as a consequence,
will not consider the latter in this
 paper. In view of the cancellation of contributions of different sign
 we also decide to include explicitly only the lowest-mass
 ground state in the mixed-parity sum rules, which differs from the
 procedure adopted by the authors of Ref.~\cite{lenz}.

Let us now turn to the question how to choose the 
Borel parameter $M^2$ and the 
continuum threshold $s_0$, the internal sum rule parameters.
As mentioned before, the dependence of the sum rules on $s_0$ is weak
and so we simply use the same values of $s_0$ as for the quark
current sum rules, i.e.\
$s_0(K) = (1.1\pm 0.3)\,{\rm GeV}^2$, $s_0^\parallel(K^*) = 
(1.7\pm 0.3)\,{\rm GeV}^2$ and $s_0^\perp(K^*) = (1.3\pm 0.3)\,
{\rm GeV}^2$ \cite{BZ}. The small dependence on $s_0$ also allows one to
use slightly higher values of $M^2$ than the usual 1 to $2\,{\rm GeV}^2$,
which improves the convergence of the operator product expansion of
the correlation functions and reduces the variation of the sum rule
with $M^2$. We choose $M^2 = (1.6 \pm 0.4)\, {\rm
  GeV^2}$ for $K$ and $M^2 = (1.8
\pm 0.4)\, {\rm GeV^2}$ for $K^*$.

After this general discussion of the choice of sum rules and
parameters let us now turn to the three $\kappa_4$ parameters in turn.

\subsection{\boldmath $\kappa_4(K)$}

The mixed-parity sum rule for $\kappa_4(K)$ is obtained from the
correlation function $\Pi^{(a)}_{G,2}$ in App.~\ref{appA},
Eq.~(\ref{piga}), and given by
\begin{eqnarray}
\lefteqn{f_K^2 m_K^2 \kappa_4(K)\, e^{-m^2_K/M^2} 
%+  (f_{K_1}^\parallel)^2 m_{K_1}^2 \kappa_4^\parallel(K_1) e^{-m^2_{K_1}/M^2}
= \frac{\alpha_s}{72 \pi^3}\,(m_s^2-m_q^2)\int_0^{s_0}ds\, e^{-s/M^2}\left(10
\ln\frac{s}{\mu^2} -25 \right)}\hspace*{2cm}\nonumber\\
&&{}+\frac{2}{9}\,\frac{\alpha_s}{\pi}
\left(m_s\quark-m_q\squark\right) \left\{ -\frac{1}{3} + \gamma_E
-\ln\frac{M^2}{\mu^2}+ \int_{s_0}^\infty
\frac{ds}{s}\,e^{-s/M^2}\right\} \nonumber\\
&&{} + \frac{10}{9}\,\frac{\alpha_s}{\pi}\left(m_s\squark -m_q\quark\right)
+\frac{1}{6M^2}\left(m_s \smixed -m_q\mixed\right)\nonumber\\
&&{}+
\frac{m_s^2-m_q^2}{6M^2}\gluon\left\{1-
\frac{1}{2}\left(\ln\frac{M^2}{\mu^2}-\gamma_E+1\right)
-M^2\int_{s_0}^\infty \frac{ds}{2 s^2}\,e^{-s/M^2}\right\}\nonumber\\
&&{}  +\frac{8\pi\alpha_s}{27 M^2}\,[\quark^2-\squark^2]\,.\label{SRk4K}
\end{eqnarray}
This sum rule includes all relevant contributions up to dimension six.
Numerically, all dominant contributions have the same sign, with the
largest one from $\squark$, followed by the ones from $\smixed$ and
perturbation theory which are roughly of the same size. 

In Fig.~\ref{fig:1} we plot the resulting values for $\kappa_4(K)$
and, via (\ref{a1K}), $a_1(K)$, displaying, for illustration, 
explicitly the dependence on
$\alpha_s$ and $\delta_{3,5}$. It is evident that the dependence of both 
quantities on $\delta_3$ and $\delta_5$ is nonnegligible; at the same
time, the comparison with $a_1(K)$ obtained in Ref.~\cite{BZ} 
from a QCD sum rule for  quark currents shows that both
sum rules agree within errors.\footnote{The results from the quark
  current sum rules quoted in this paper are slightly larger than
  the ones given in Ref.~\cite{BZ}. This is due to the fact that
  we have included infrared sensitive terms of type
  $m_s^2\ln(M^2/m_s^2)$ in the
  contribution of the gluon condensate  in the
  mixed quark-quark-gluon condensate rather than in the
  Wilson-coefficient of the gluon condensate, cf.\ the discussion in
  App.~\ref{appA} and Ref.~\cite{logms}.} Note that the inclusion of the
perturbative contribution is crucial: without it, we would have
obtained a {\em negative} result for $a_1(K)$. The impact of nonzero
$m_q$ is also relevant and shifts the central value of $a_1(K)$ by
$+0.025$.
 
\begin{figure}[tb]
$$\epsfxsize=0.47\textwidth\epsffile{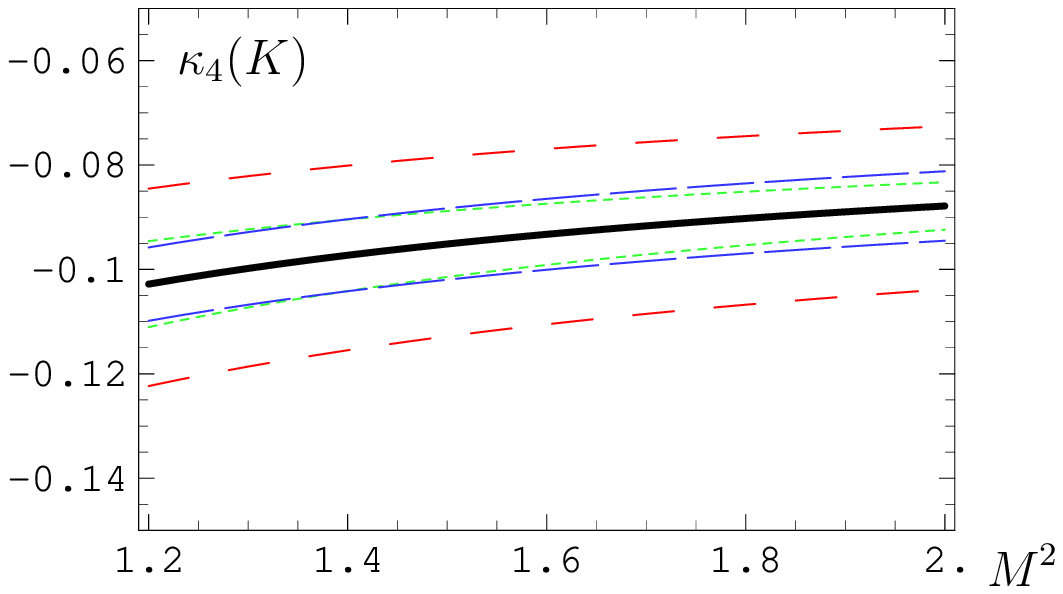}\quad
\epsfxsize=0.47\textwidth\epsffile{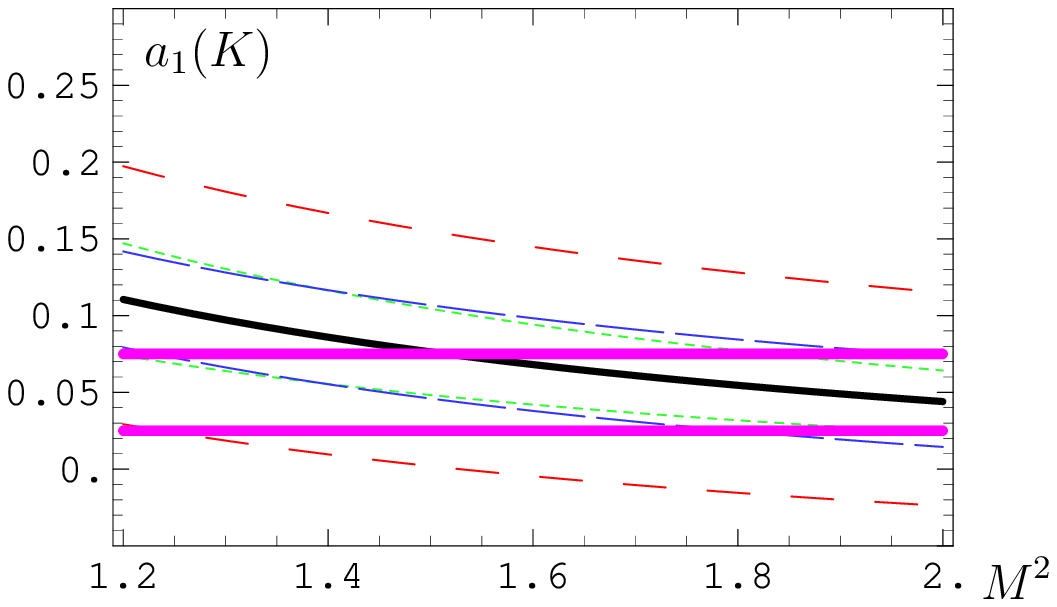}$$
\vspace*{-30pt}
\caption[]{(Colour online) 
Left panel: $\kappa_4(K)$ from (\ref{SRk4K}) as function
  of the Borel parameter $M^2$. Parameters: renormalisation scale 
$\mu=1\,$GeV, $s_0 = 1.1\,{\rm GeV}^2$.
  Solid line: central value of input parameters; dashed lines: red:
  $\delta_3=0,0.4$, green: $\delta_5=0,0.4$, blue: 
$\alpha_s(m_Z) = 0.1167,0.1207$. Right panel: $a_1(K)$ as function of $M^2$ 
from the operator relation
  (\ref{a1K}) (colour-coded as in the left panel) and the value of $a_1(K)$ 
determined in Ref.~\cite{BZ} (purple lines).}\label{fig:1}
$$\epsfxsize=0.47\textwidth\epsffile{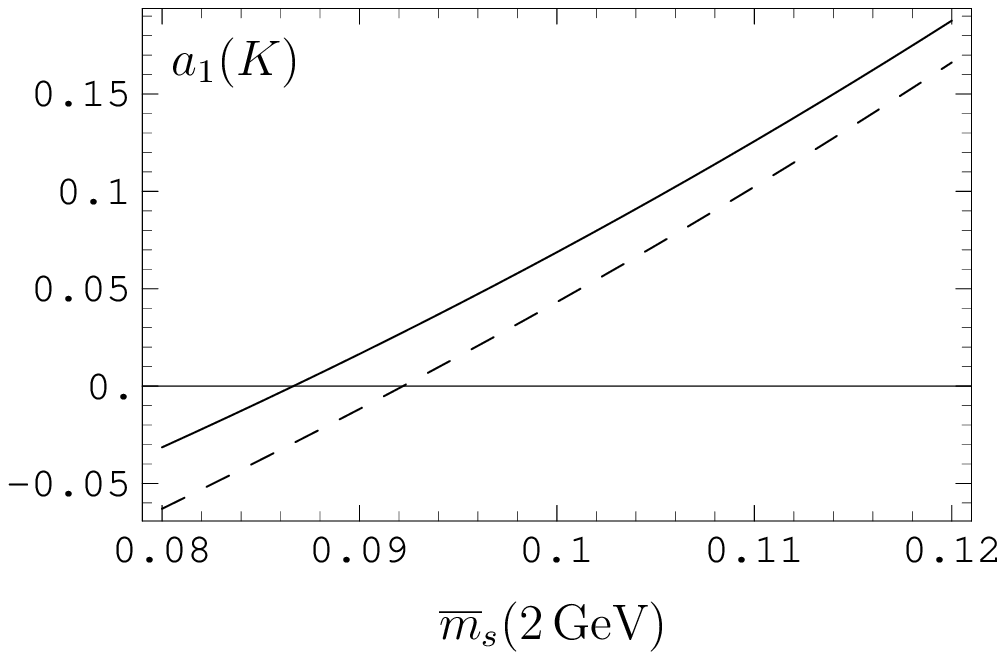}$$
\vspace*{-30pt}
\caption[]{Dependence of the central value of $a_1(K)$ from
  (\ref{a1K}) on $\overline{m}_s(2\,{\rm GeV})$. Solid line:
  $\overline{m}_q(2\,{\rm GeV}) = 4\,$MeV, dashed line:
$\overline{m}_q(2\,{\rm GeV}) = 0\,$MeV.}\label{fig:2}
\end{figure}

As for the theoretical uncertainties of $\kappa_4(K)$ and $a_1(K)$ we
note that they arise first from the QCD sum rule parameters and second
from the uncertainties of the hadronic parameters given in
Tab.~\ref{tab:1}. As for the former, as  already stated above, we
choose $M^2 = (1.6\pm 0.4)\,{\rm GeV}^2$ and $s_0 = (1.1\pm 0.3)\,{\rm
  GeV}^2$ and add the corresponding uncertainties in quadrature. 
As for the latter, we treat $m_q$, $m_s$, $\quark$, $\mixed$,
$\delta_3$ and $\delta_5$ as 
parameters with correlated errors. Chiral
perturbation theory helps to unravel some of these correlations: 
for instance, one has $(m_s+m_q)/(2m_q) = m_K^2/m_\pi^2$ and
$m_K^2  = -2(m_s+m_q)\quark/f_\pi^2$
in LO chiral perturbation theory \cite{chPT}. 
The dependence of $\delta_{3,5}$ on
$m_s$ is unfortunately unknown (and indeed would deserve further
study). In order to estimate the uncertainty
of $\kappa_4(K)$ and $a_1(K)$, we hence eliminate, using the above
relations, $m_q$ and $\quark$ as independent parameters in favour of $m_s$, 
but keep $m_0^2=\mixed/\quark$ and $\delta_{3,5}$. This procedure is likely to
overestimate the uncertainties induced by $\squark$ and $\smixed$,
but it is difficult to do better at present.
Varying all remaining independent input parameters within their respective
ranges given in Tab.~\ref{tab:1}, we obtain the
following results: 
\begin{eqnarray}
\kappa_4(K) & = & -0.09\pm 0.01 \pm 0.01\pm 0.01\pm 0.02\pm 0.01\pm
0.00 = -0.09\pm 0.01\pm 0.02,\nonumber\\
a_1(K) & = & \phantom{-}0.07 \pm 0.04 \pm 0.03 \pm 0.11 \pm 0.07 \pm 0.03 \pm
0.01 =\phantom{-} 0.07 \pm 0.04\pm0.14,\label{36a}
\end{eqnarray}
where the first uncertainty comes from the variation of the sum rule
specific parameters $M^2$ and $s_0$, the second one from $\alpha_s$,
the 3rd from $m_s$,
the 4th from $\delta_3$, the 5th from $\delta_5$
and the 6th from $m_0^2 = \mixed/\quark$. For the total
uncertainty we give two terms: the first comes from the sum rule
parameters and the second is obtained by adding all hadronic 
uncertainties in quadrature. As mentioned before, any
uncertainty of $\kappa_4(K)$ induces a corresponding uncertainty in
$a_1(K)$ that is about four times larger, except for the strange quark
masses whose uncertainty also plays in the second term on the
right-hand side of (\ref{a1K}). The dependence of $a_1(K)$ on $m_s$ is
shown in Fig.~\ref{fig:2}. The effect of nonzero $m_q$ in the first term
on the right-hand side of (\ref{a1K}) is a shift by $+0.04$, which is 
partially, but not completely, compensated by the
$m_q$-dependent contributions to $\kappa_4(K)$.
Comparing with the value of $a_1(K)$
quoted in Ref.~\cite{lenz}, Eq.~(\ref{reslenz}), we see that the
central value in (\ref{36a}) 
is smaller and also the total uncertainty is larger. 
The larger error is due
to the fact that we have chosen slightly larger errors for $m_s$ and
also have included the uncertainty induced by $\alpha_s$. 

Let us now
compare the result (\ref{36a}) with the one obtained from quark
current sum rules \cite{BZ}, with the same sequence of errors as in
(\ref{36a}):
\begin{equation}
a_1(K)^{\rm BZ} = 0.06\pm 0.01\pm0.00\pm0.01\pm0.01\pm0.01\pm0.00 =
0.06\pm 0.01\pm 0.02\,.
\end{equation}
This number is slightly larger than the one quoted in Ref.~\cite{BZ},
cf.\ footnote~2. Although the central values of $a_1(K)$ agree very
well and hence confirm the consistency of the sum rule results, 
it is obvious that  the operator relation (\ref{a1K}) cannot
match the accuracy of the quark current sum rule and is hence not very
useful for constraining $a_1(K)$. 

\subsection{\boldmath $\kappa_4^\parallel(K^*)$}

Let us now turn to $\kappa_4^\parallel(K^*)$. The mixed-parity sum
rule is derived from the correlation function $\Pi_{G,2}^{(v)}$ in
App.~\ref{appA}, Eq.~(\ref{pigv}), and reads
$$
\kappa_4^\parallel(K^*) (f^\parallel_K)^2 m_{K^*}^2 
e^{-m^2_{K^*}/M^2} = 
(m_s^2-m_q^2)\,\frac{\alpha_s}{72 \pi^3}\,\int_0^{s_0}ds\, e^{-s/M^2}\left(10
\ln\frac{s}{\mu^2} -25 \right)
$$
\begin{eqnarray}
&&{}-\frac{2\alpha_s}{9 \pi}\,(m_s\quark-m_q\squark)
\left\{ -\frac{1}{3} + \gamma_E
-\ln\frac{M^2}{\mu^2}+ \int_{s_0}^\infty
\frac{ds}{s}\,e^{-s/M^2}\right\}\nonumber\\
&&{} + 
\frac{10\alpha_s}{9 \pi}\,(m_s\squark-m_q\quark)
+\frac{1}{6M^2}\,(m_s \smixed -m_q\mixed)\nonumber\\
&&{}+\frac{m_s^2-m_q^2}{6M^2}\gluon\left\{1-
\frac{1}{2}\left(\ln\frac{M^2}{\mu^2}-\gamma_E+1\right)
-M^2\int_{s_0}^\infty \frac{ds}{2 s^2}\,e^{-s/M^2}\right\}\nonumber\\
&&{}  +\frac{8\pi\alpha_s}{27 M^2}[\quark^2-\squark^2]\,.\label{36}
\end{eqnarray}
The resulting values of $\kappa_4^\parallel(K^*)$ and
$a_1^\parallel(K^*)$ are shown in Fig.~\ref{fig:3}. 
\begin{figure}
$$\epsfxsize=0.47\textwidth\epsffile{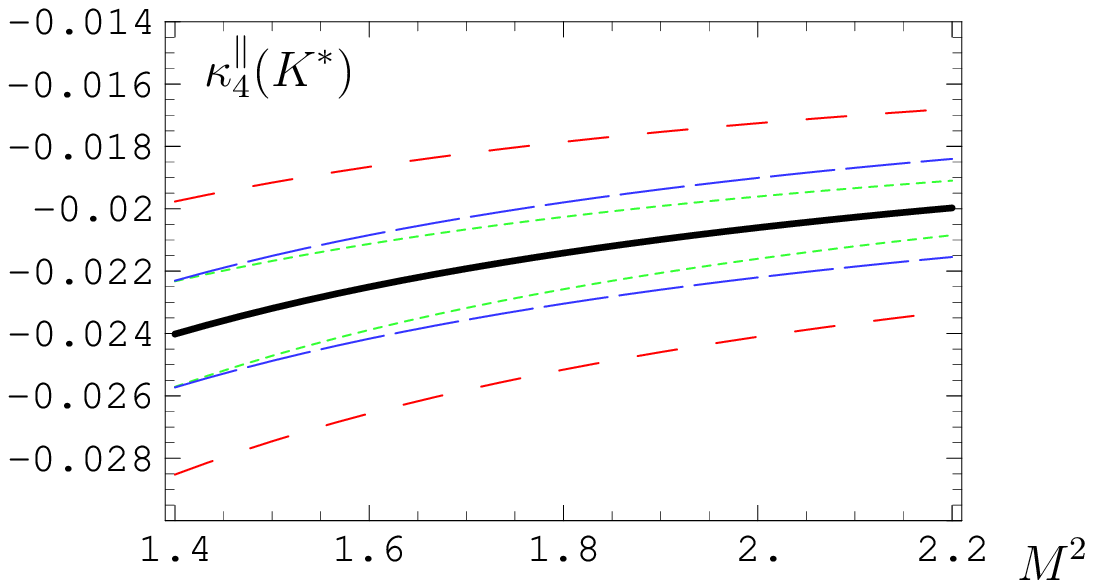}\quad
\epsfxsize=0.47\textwidth\epsffile{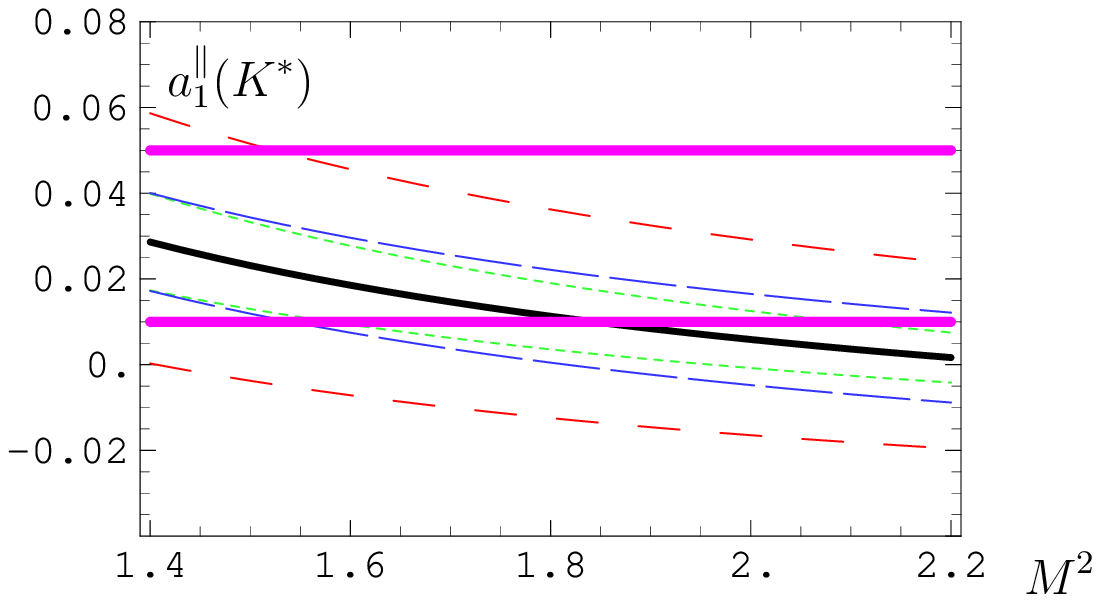}$$
\vspace*{-30pt}
\caption[]{Left panel: $\kappa_4^\parallel(K^*)$ from (\ref{36}) as function
  of the Borel parameter $M^2$. Parameters: renormalisation scale 
$\mu=1\,$GeV, $s_0 = 1.7\,{\rm GeV}^2$. Solid black line: central
  values of parameters; the coloured lines have the same meaning as in
  Fig.~\ref{fig:1}. Right panel: 
$a_1^\parallel(K^*)$ as function of $M^2$ from the operator relation
  (\ref{a1Kpar}) and the  sum rule for $a_1^\parallel(K^*)$ 
calculated in Ref.~\cite{BZ} (purple lines).  }\label{fig:3}
\end{figure}
Again,
the contribution from perturbation theory is crucial numerically:
without it, the resulting values of $a_1^\parallel(K^*)$ would have
been negative. Our final results are:
\begin{eqnarray}
\kappa_4^\parallel(K^*) & = & -0.022\pm 0.003\pm 0.001\pm 0.003\pm
0.004 \pm 0.001\pm 0.001\nonumber\\
& = &-0.022\pm 0.003\pm0.005,\nonumber\\
a_1^\parallel(K^*) & = & \phantom{-}0.01\pm 0.02 \pm 0.01 \pm 0.01 \pm
0.02 \pm 0.01 \pm 0.00\nonumber\\
& = & \phantom{-}0.01 \pm 0.02\pm 0.03\label{41a}
\end{eqnarray}
with the same assignment and treatment of uncertainties as in (\ref{36a}); the
uncertainty coming from $f_K^\perp$ is included in that from
$m_s$. In contrast to the pseudoscalar case, the translation of 
$\kappa_4^\parallel(K^*)$ into $a_1^\parallel(K^*)$ does not increase
the uncertainty from $m_{s}$ any more than the other uncertainties,
so that the total error of $a_1^\parallel(K^*)$ is smaller than
that of $a_1(K)$. The impact of $m_q$-dependent terms in negligible. 
The results (\ref{41a}) differ from those of
Ref.~\cite{lenz}, (\ref{28}) and (\ref{29}), where the pure-parity sum
rule has been used instead. 
The result from the quark current sum rule is
\begin{equation}
a_1^\parallel(K^*)^{\rm BZ}= 0.03\pm 0.02.
\end{equation}
Again we
find agreement between the results for $a_1$ from the sum rules for
$\kappa_4$ and the quark current sum rules, but at the same time the
uncertainty of the former is larger than that of the latter.

\subsection{\boldmath $\kappa_4^\perp(K^*)$}

The last parameter left to be determined is
$\kappa_4^\perp(K^*)$. Its mixed-parity sum rule is derived from the
correlation function $\Pi_{G,4}$, Eq.~(\ref{pig4}), and reads
\begin{eqnarray}
\lefteqn{\kappa_4^\perp(K^*)(f_{K}^\perp)^2 m_{K^*}^2 
e^{-m_{K^*}^2/M^2} = (m_s^2-m_q^2)\,\frac{\alpha_s}{72 \pi^3} \int_0^{s_0}
ds\,e^{-s/M^2} \left(-6 \ln\frac{s}{\mu^2} + 14 \right)}\nonumber\\
&&{}+\frac{m_s \alpha_s}{3 \pi}\! \left\{ \frac{1}{3}\,\quark - 2
\squark\right\} - \frac{m_q \alpha_s}{3 \pi}\! \left\{ \frac{1}{3}\,\squark - 2
\quark\right\}+\frac{1}{6 M^2}\left(m_q\mixed - m_s\smixed\right)\nonumber\\
&&{}+\frac{m_s^2-m_q^2}{12M^2}\gluon\left\{-2 + 
\left((\ln\frac{M^2}{\mu^2}-\gamma_E+1\right)
+ M^2\!\int_{s_0}^\infty
\frac{ds}{s^2}\,e^{-s/M^2}\right\}.\label{43a}
\end{eqnarray}
The results for $\kappa_4^\perp(K^*)$ and $a_1^\perp(K^*)$ are shown
in Fig.~\ref{fig:4};
\begin{figure}
$$\epsfxsize=0.45\textwidth\epsffile{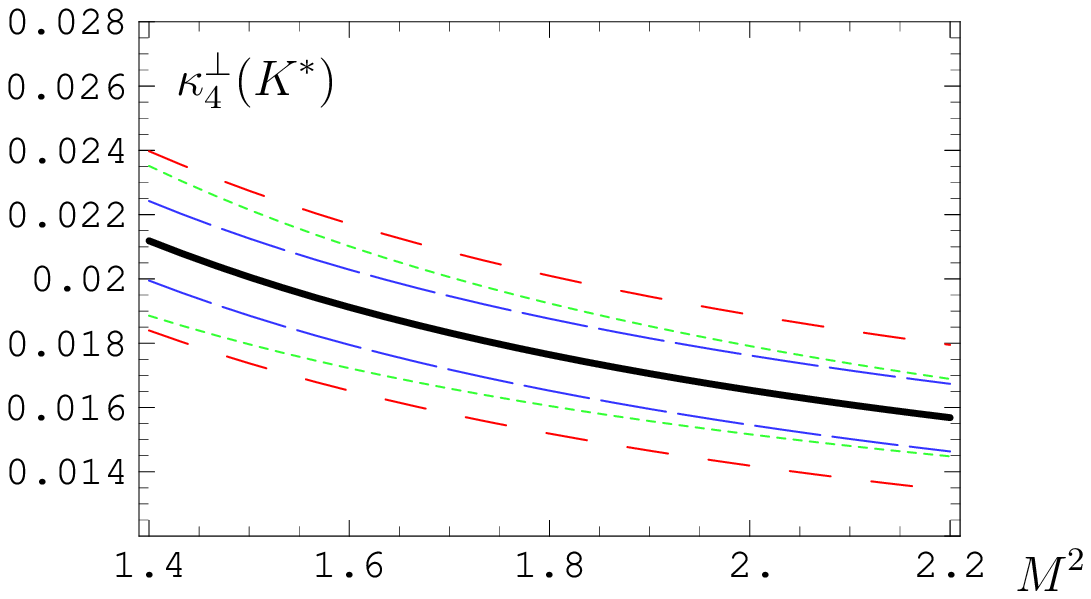}\quad
\epsfxsize=0.45\textwidth\epsffile{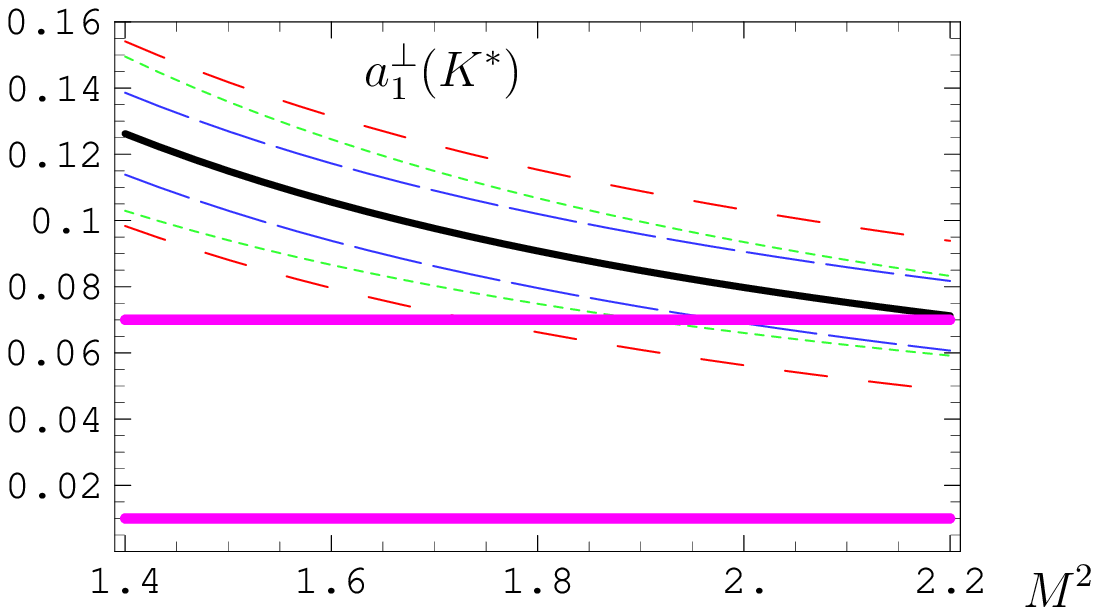}$$
\vspace*{-20pt}
\caption[]{Left panel: $\kappa_4^\perp(K^*)$ from (\ref{43a}) as function
  of the Borel parameter $M^2$. Parameters: renormalisation scale 
$\mu=1\,$GeV, $s_0 = 1.3\,{\rm GeV}^2$. Solid black line: central
  values of parameters; the coloured lines have the same meaning as in
  Fig.~\ref{fig:1}. Right panel: $a_1^\perp(K^*)$ as function of $M^2$ 
from the operator relation
  (\ref{eq:perp}) and the  sum rule for $a_1^\perp(K^*)$ 
calculated in Ref.~\cite{BZ} (purple lines).}\label{fig:4}
\end{figure}
including uncertainties, we find
\begin{eqnarray}
\kappa_4^\perp(K^*) & = & 0.018\pm 0.004\pm 0.001\pm 0.002\pm 0.002
\pm 0.002 \pm 0.001\nonumber\\
& = &0.018\pm 0.004\pm 0.004\,,\nonumber\\
a_1^\perp(K^*) & = & 0.09\pm 0.04 \pm 0.01\pm 0.01\pm 0.02\pm 0.02\pm
0.01\nonumber\\
&=& 0.09\pm 0.04\pm 0.03\,.\label{43}
\end{eqnarray}
Note that the ``enhancement'' factor of uncertainties of
$a_1^\perp(K^*)$ due to
$\kappa_4^\perp(K^*)$ is 10, which is the reason for the large total
uncertainty in (\ref{43}). The impact of $m_q$-dependent terms is
again negligible. The quark current sum rule yields \cite{BZ} 
\begin{equation}
a_1^\perp(K^*)^{\rm BZ} = 0.04\pm 0.01 \pm 0.01\pm 0.01 \pm 0.01 \pm 0.00 \pm
0.00 = 0.04\pm 0.01 \pm 0.02.
\end{equation}
Hence, also for $a_1^\perp(K^*)$ do the results of the two approaches
agree within errors, with the quark current sum rule being more accurate.

\section{Summary and Conclusions}\label{sec:4}

In this paper, we have obtained the following relations for the first 
Gegenbauer
moments of the leading-twist distribution amplitudes of $K$ and $K^*$ mesons:
\begin{eqnarray}
\frac{9}{5}\, a_1(K) &=& -\frac{m_s-m_q}{m_s+m_q} +
4\,\frac{m_s^2-m_q^2}{m_{K}^2}  - 8 \kappa_{4}(K),\nonumber\\
\frac{3}{5}\,a_1^\parallel(K^*) &=&
-\frac{f_K^\perp}{f_K^\parallel}\,\frac{m_s-m_q}{m_{K^*}} + 2 
\,\frac{m_s^2-m_q^2}{m_{K^*}^2} - 4 \kappa^{\parallel}_{4}(K^*),\nonumber\\
\frac{3}{5}\, a_{1}^\perp(K^*) &=&
-\frac{f_K^\parallel}{f_K^\perp}\,\frac{m_s-m_q}{2m_{K^*}} +
  \frac{3}{2}\,\frac{m_s^2-m_q^2}{m_{K^*}^2} + 6 \kappa^{\perp}_4(K^*),
\label{40}
\end{eqnarray}
where the $\kappa_4$ matrix elements are defined as
\begin{eqnarray*}
\langle 0 | \bar q  
(g G_{\alpha\mu}) 
i\gamma^\mu\gamma_5  s|K(q)\rangle &=& i q_\alpha f_{K} m_K^2 
\kappa_{4}(K)\,,\\
\langle 0 | \bar q (g G_{\alpha\mu}) i\gamma^\mu   s | K^*(q,\lambda) \rangle
&=& e^{(\lambda)}_\alpha  f_K^\parallel m_{K^*}^3
\kappa^{\parallel}_{4}(K^*)\,,\nonumber\\[-40pt]
\end{eqnarray*}
\begin{eqnarray*} 
\lefteqn{
\langle 0 | \bar q (g G_{\alpha}^{\phantom{\alpha}\mu}) \sigma_{\beta\mu} s |
K^*(q,\lambda)\rangle=}\hspace*{1.5cm}\nonumber\\ 
&=& f_K^\perp m_{K^*}^2 \left\{ \frac{1}{2}\,
\kappa^\perp_3(K^*) (e^{(\lambda)}_\alpha q_\beta + 
e^{(\lambda)}_\beta q_\alpha) + \kappa_4^{\perp}(K^*)
(e^{(\lambda)}_\alpha q_\beta - e^{(\lambda)}_\beta q_\alpha)\right\} .
\end{eqnarray*}
The first two relations in (\ref{40}) were already derived in
Ref.~\cite{lenz}, the third is new. We have interpreted these
relations as constraints on $a_1$ and calculated the three $\kappa_4$
parameters 
from QCD sum rules. We have improved the sum rules given in
Ref.~\cite{lenz} for $\kappa_4(K)$ and $\kappa_4^\parallel(K^*)$ by including 
two-loop perturbative contributions, the gluon condensate
contribution and terms in $m_q$; 
the former  proved to be very relevant numerically, 
the terms in $m_q$ are relevant for $a_1(K)$. 
We have also derived a new sum rule for $\kappa_4^\perp(K^*)$ to the same
accuracy. All these sum rules exhibit only a small continuum
contribution and all relevant contributions come with equal sign.
The results for $a_1$ obtained from the relations (\ref{40}) agree,
within errors, with those obtained in Ref.~\cite{BZ} from quark
current sum rules which is an important confirmation of the
consistency of QCD sum rule calculations of these quantities and 
strengthens our confidence in the results. From a phenomenological
point of view, however, the operator 
relations (\ref{40}) are, at least at present,
less useful than the quark current sum rules for $a_1$, as the
uncertainties of the $\kappa_4$ parameters are too large to allow an
accurate determination of $a_1$. The uncertainties of $\kappa_4$
arise from (a) the dependence of the sum rule on the sum rule internal
parameters $M^2$ and $s_0$, (b) the uncertainties of $\alpha_s$ at the 
hadronic scale
$\sim\,$1~GeV and (c) the uncertainties of $m_s$ and the SU(3) breaking
of quark and mixed condensates parametrised by $\delta_{3,5}$.
All these uncertainties
enter $a_1$ multiplied by large factors 5 to 10, Eqs.~(\ref{40}). In
contrast, the quark current sum rules for $a_1$ studied in
Refs.~\cite{alex,BZ} are not very sensitive to these effects and come
with smaller uncertainties. We hence suggest that the
relations (\ref{40}) be interpreted as constraints on $\kappa_4$ rather than
$a_1$. Using the updated values of $a_1$ from quark current sum rules
quoted in Sec.~\ref{sec:3}, adding the errors linearly,
\begin{equation}\label{xyz}
a_1(K)^{\rm BZ} = 0.06\pm 0.03, \quad a_1^\parallel(K^*)^{\rm BZ} = 
0.03\pm 0.02\quad 
a_1^\perp(K^*)^{\rm BZ} = 0.04\pm 0.03,
\end{equation}
we find by solving (\ref{40}) for $\kappa_4$:
\begin{eqnarray*}
\kappa_4(K) & = & -\frac{1}{8}\,\frac{m_s-m_q}{m_s+m_q} -
\frac{9}{40}\,a_1(K) + \frac{m_s^2-m_q^2}{2m_K^2} = -0.09\pm 0.02,\nonumber\\
\kappa^\parallel_4(K^*) & = &
-\frac{f_K^\perp}{f_K^\parallel}\,\frac{m_s-m_q}{4 m_{K^*}} -
\frac{3}{20}\,a_1^\parallel(K^*) + \frac{m_s^2-m_q^2}{2 m_{K^*}^2}
= -0.024 \pm 0.003,\nonumber\\
\kappa^\perp_4(K^*) & = &
\frac{f_K^\parallel}{f_K^\perp}\,\frac{m_s-m_q}{12 m_{K^*}} +
\frac{1}{10}\,a_1^\perp(K^*) - \frac{m_s^2-m_q^2}{4 m_{K^*}^2}
= 0.012\pm 0.004.
\end{eqnarray*}
For $\kappa_4(K)$ and $\kappa_4^\parallel(K^*)$ 
the central value agrees well with the results from the direct calculation, for
$\kappa_4^{\perp}(K^*)$ 
there is agreement within errors.
How can these results be improved? The quark
current results for
$a_1$ would profit from a calculation of perturbative radiative
corrections $\sim m_s^2\alpha_s$, which is technically feasible, but
beyond the scope of this paper. Both $a_1$ and $\kappa_4$ would benefit
from a reduction of the errors of $m_s$.

In summary, we hope that the present paper helps to settle the
controversy about $a_1$ which started from the observation that
the original calculation of Ref.~\cite{Russians} suffers from  a
sign-mistake of the perturbative contribution, which was corrected in
Ref.~\cite{elena}. Unfortunately, the chiral-odd sum rules used in
Ref.~\cite{elena} come with large
cancellations of the dominant contributions and are hence not very
useful for precise calculations of $a_1$. In Ref.~\cite{alex},
$a_1(K)$ was then determined from chiral-even quark current sum rules and
in Ref.~\cite{BZ} also $a_1^{(\perp,\parallel)}(K^*)$ was calculated
using that method. These sum rules do not exhibit any cancellations of
large contributions and are stable under the variation of all
input parameters. As we have shown in this paper, these results agree with
those from the operator relations
(\ref{40}) within errors, but are 
more accurate. We conclude that the quark current sum rule results (\ref{xyz})
present the presently best determination of $a_1$. Given the
phenomenological importance of $a_1$, an independent calculation on
the lattice would be both timely and useful and we would like to appeal
to the lattice community to take up the challenge.

\subsection*{Acknowledgements}
R.Z.\ is supported in part by the EU-RTN Programme, Contract No.\
HPEN-CT-2002-00311, ``EURIDICE''. 

\appendix
\setcounter{equation}{0}
\renewcommand{\theequation}{A.\arabic{equation}}

\section{Correlation Functions}\label{appA}

In this appendix we give the relevant formulas for the correlation
functions from which the QCD sum rules given in Sec.~\ref{sec:3} are obtained.
The correlation functions are of the generic form
\begin{equation}
\label{eq:master}
\Pi_{\alpha\dots}(q) = 
i\int d^4y e^{iqy} \langle 0 | T [\bar q (g
  G_{\alpha\mu})\Gamma_1^\mu  s](y) 
[\bar s \Gamma_2 q](0)|0\rangle\,,
\end{equation}
where $\Gamma_1^\mu$ and $\Gamma_2$ are suitably chosen Dirac
structures. The dots stand for additional indices from 
$\Gamma_2$.

\subsection{\boldmath $\kappa_4(K)$}

$\kappa_4(K)$ can be extracted from either a pure-partity sum rule, to
which only pseudoscalar states contribute, or a mixed-parity sum rule
which also contains contributions from axialvector mesons. 
As for pure-parity sum rules, 
one possible choice of the Dirac structures is  $\Gamma_1^\mu
= i\gamma^\mu\gamma_5$ and $\Gamma_2=i\gamma_5$, which results in the
correlation function
\begin{equation}
\Pi_\alpha(q) = i q_\alpha \Pi_G^{(p)}(q^2).
\end{equation}
Another choice is $\Gamma_1^\mu
= i\gamma^\mu\gamma_5$ as before and $\Gamma_2=\gamma_\beta\gamma_5$,
with the correlation function
\begin{equation}
\Pi_{\alpha\beta}(q)  = g_{\alpha\beta} \Pi^{(a)}_{G,1}(q^2) +
q_\alpha q_\beta  \Pi^{(a)}_{G,2}(q^2),  
\end{equation}
where $\Pi^{(a)}_{G,1}$ receives contributions from $1^+$ intermediate
states only, whereas $\Pi^{(a)}_{G,2}$ is a mixed-parity correlation
function with contributions from both $0^-$ and $1^+$ states.

These three correlation functions are not independent of each other,
but related by $\partial^\beta \bar s \gamma_\beta\gamma_5 q  = (m_s+m_q)
\bar s i\gamma_5 q$, so that
\begin{equation}\label{scalarEOM}
\Pi^{(a)}_{G,1}(q^2) + q^2 \Pi^{(a)}_{G,2}(q^2) 
= (m_s+m_q)\Pi_G^{(p)} + \mbox{ contact terms},
\end{equation}
where the contact terms are independent of $q^2$.
As terms in $m_q$ are numerically relevant in the operator relation
(\ref{a1K}), we calculate the  correlation functions to $O(m_q)$ and find
\begin{eqnarray}
 \Pi_{G}^{(p)}(q^2) &=& 
-(m_s-m_q)\,\frac{\alpha_s}{48\pi^3} q^2\left[\ln^2\frac{-q^2}{\mu^2}-
\ln\frac{-q^2}{\mu^2}\right]
-\frac{1}{4q^2}\left[\mixed  -\smixed\right]
\nonumber\\&&{}
 -\frac{\alpha_s }{3\pi} \left[ \quark - \squark \right] 
\ln\frac{-q^2}{\mu^2}\nonumber\\
&&{}     +\frac{1}{8 q^2}  \gluon 
       \left[ m_s \left(1-\ln \frac{-q^2}{m_s^2}\right) -
m_q \left(1-\ln\frac{-q^2}{m_q^2}\right)\right],\nonumber\\
\Pi_{G,1}^{(a)}(q^2) &=&  
(m_s^2-m_q^2)\,\frac{\alpha_s}{144\pi^3} q^2\left[7\ln^2\frac{-q^2}{\mu^2}-
47\ln\frac{-q^2}{\mu^2}\right]\nonumber\\
&&-\frac{m_s\alpha_s}{3\pi}
   \left[ \frac{5}{3} \quark - \squark \right]\ln\frac{-q^2}{\mu^2}
+ \frac{m_q\alpha_s}{3\pi}
   \left[ \frac{5}{3} \squark - \quark
     \right]\ln\frac{-q^2}{\mu^2}\nonumber\\
&& - \left(\frac{m_q}{12} + \frac{m_s}{4}\right)
\frac{\mixed}{q^2} 
  + \left(\frac{m_s}{12}+\frac{m_q}{4}\right)\frac{\smixed}{q^2}
+ \frac{m_q m_s}{8q^2}\,\gluon\,\ln\frac{m_s^2}{m_q^2}\nonumber
\end{eqnarray}
\begin{eqnarray}
&&{}-\frac{m_s^2}{24 q^2}  \gluon 
       \left[ 1 + \ln\frac{-q^2}{m_s^2}
         \right] +\frac{m_q^2}{24 q^2}  \gluon 
       \left[ 1 + \ln\frac{-q^2}{m_q^2}\right]\nonumber\\
&&{}-\frac{8\pi\alpha_s}{27 q^2}\big[\quark^2 -\squark^2\big],
\end{eqnarray}
\begin{eqnarray}
\Pi_{G,2}^{(a)}(q^2) &=&  
(m_s^2-m_q^2)\,\frac{\alpha_s}{72\pi^3} \left[-5\ln^2\frac{-q^2}{\mu^2}+
25\ln\frac{-q^2}{\mu^2}\right]\nonumber\\
&&{}+\frac{2m_s\alpha_s}{9\pi q^2}\quark
   \left[\frac{1}{3} + \ln\frac{-q^2}{\mu^2}\right]
  -\frac{10m_s\alpha_s}{9\pi q^2}\squark
\nonumber\\&&{} 
-\frac{2m_q\alpha_s}{9\pi q^2}\squark
   \left[\frac{1}{3} + \ln\frac{-q^2}{\mu^2}\right]
  +\frac{10m_q\alpha_s}{9\pi q^2}\quark
+ \frac{m_s}{6q^4} \smixed - \frac{m_q}{6q^4} \mixed\nonumber\\
&&{} +\frac{m_s^2}{6 q^4}  \gluon 
       \left[ 1 -\frac{1}{2}\, \ln\frac{-q^2}{m_s^2} \right]-
\frac{m_q^2}{6 q^4}  \gluon 
\left[ 1 -\frac{1}{2}\, \ln\frac{-q^2}{m_q^2} \right]
\nonumber\\&&{}
+\frac{8\pi\alpha_s}{27 q^4}\big[\quark^2 -\squark^2\big].
\label{piga}
\end{eqnarray}
The expression for $\Pi_{G}^{(p)}$ has already been given in
Ref.~\cite{lenz}, together with $\Pi_{G,(1,2)}^{(a)}$, to leading
order in SU(3) breaking. The terms in $m_s^2$ and $m_q$ are new. The above
expressions fulfill the relation (\ref{scalarEOM}). 

At this point a few comments are in order concerning the structure of
these formulas. The reader may have noticed that the Wilson
coefficient of the gluon condensate contributions to the above
correlation functions contain infrared sensitive terms $\sim\ln
(-q^2/m_{q,s}^2)$. These terms appear to violate the structure of the
operator product expansion which stipulates that long- and
short-distance contributions be properly factorised and all long-distance
contributions be absorbed into the condensates, leaving purely
short-distance Wilson coefficients which must be analytic
in $m_{q,s}$. As discussed in Ref.~\cite{logms}, the appearance of terms
logarithmic in $m_{q,s}$ is due to the fact that the above expressions are
obtained using Wick's theorem to calculate the condensate
contributions, which implies that the condensates are normal-ordered: 
$\langle O
\rangle = \langle 0 |\!:\!O\!:\! |0\rangle$. Recasting the OPE in terms of
non-normal-ordered operators, all infrared sensitive terms can be
absorbed into the corresponding condensates. Indeed, using \cite{logms}
$$
\langle 0 | \bar s g G s | 0 \rangle = \langle 0| \!:\!\bar s g G
s\!:\!| 0 \rangle +
\frac{m_s}{2}\, \log\,\frac{m_s^2}{\mu^2} \,\langle 0 |\! :\!
\frac{\alpha_s}{\pi}\, G^2\!:\!| 0\rangle,
$$
and the corresponding formula for $q$ quarks,
all terms $\sim \ln m_{q,s}^2$ can be absorbed into the mixed
quark-quark-gluon condensate and the resulting Wilson-coefficients
can be expanded in powers of $m_{q,s}^2$. In calculating the sum rules, we
hence will use
$$
\ln\,\frac{-q^2}{m_{q,s}^2}\to \ln \frac{-q^2}{\mu^2}\,.
$$
As for the structure of the ultraviolet logarithms $\sim\ln
(-q^2/\mu^2)$, they follow from the mixing of the gluonic operator 
$\bar q (g G_{\alpha\mu}) i \gamma^\mu\gamma_5 s$ with various
quark-bilinear operators as given in Eq.~(20) in Ref.~\cite{lenz}.

\subsection{\boldmath $\kappa_4^\parallel(K^*)$}

The correlation functions used to determine $\kappa_4^\parallel(K^*)$
are very similar to those in the previous subsection. We choose
$\Gamma_1^\mu = i\gamma^\mu$ and $\Gamma_2 = \sigma_{\beta\gamma}$ to
obtain the pure-parity correlation function
\begin{equation}
\Pi_{\alpha\beta\gamma}(q)  = 
i(g_{\alpha\beta}q_\gamma-g_{\alpha\gamma}q_\beta)\Pi^{(\sigma)}_{G}(q^2) 
\end{equation}
and $\Gamma_2 = \gamma_\beta$ which yields
\begin{equation}
\Pi_{\alpha\beta}(q)  = g_{\alpha\beta} \Pi^{(v)}_{G,1}(q^2) + 
q_\alpha q_\beta  \Pi^{(v)}_{G,2}(q^2).
\end{equation}
$\Pi^{(\sigma)}_{G}$ and $\Pi^{(v)}_{G,1}$ receive 
contributions from $1^-$ states only
and $\Pi^{(v)}_{G,2}$ from both $1^-$ and $0^+$ states.
Another possible choice is $\Gamma_2 = {\mathds
  1}$ which yields the pure-parity correlation function
\begin{equation}
\Pi_\alpha(q) = q_\alpha \Pi_G^{(s)}(q^2)
\end{equation}
with contributions from only $0^+$ states. $\Pi_G^{(s)}$ and $
\Pi^{(v)}_{G,1(2)}$ are related by the equation of motion for the
vector current:
\begin{equation}
\Pi^{(v)}_{G,1}(q^2) + q^2 \Pi^{(v)}_{G,2}(q^2) 
= (m_s-m_q)\Pi_G^{(s)} + \mbox{ contact terms}.
\end{equation}
The expression for $\Pi^{(\sigma)}_{G}$ was given in
Ref.~\cite{lenz}, the other correlation functions are obtained by the
simple replacements
\begin{eqnarray}
\Pi_{G,1(2)}^{(v)}(q^2) &=&
\left.\Pi_{G,1(2)}^{(a)}(q^2)\right|_{m_q\to -m_q,\,\quark \to 
-\quark,\,\mixed\to-\mixed}  \label{pigv} \\
\Pi_{G}^{(s)}(q^2) &=& \left.\Pi_{G}^{(p)}(q^2)\right|_{m_q\to -m_q,\,
\quark \to -\quark,\,\mixed\to-\mixed}  
\end{eqnarray}
which follows from the chiral structure of the correlation functions.

\subsection{\boldmath $\kappa_4^\perp(K^*)$}

For $\kappa_4^\perp(K^*)$, $\Gamma_1^\mu$ is given by 
$\sigma^{\beta\mu}$ and for $\Gamma_2$ we choose
$\sigma_{\gamma\delta}$. The
resulting correlation function has contributions from both $1^-$
and $1^+$ states and can be written as
\begin{equation}
\Pi_{\alpha\beta\gamma\delta}(q) =  
i \Pi^{1^-}_{G,4}(q^2) P^{1^-}_{4,\alpha\beta\gamma\delta} + 
i \Pi^{1^-}_{G,3}(q^2) P^{1^-}_{3,\alpha\beta\gamma\delta} + 
i \Pi^{1^+}_{G,4}(q^2)P^{1^+}_{4,\alpha\beta\gamma\delta},
\end{equation}
where the projectors $P^{1^\pm}$ are given by 
\begin{eqnarray*}
P^{1^-}_{4,\alpha\beta\gamma\delta}&=& 
\frac{1}{q^2}\,\left[(g_{\alpha\gamma} q_\beta q_\delta - \{ \alpha 
\leftrightarrow \beta \}) - 
\left(\{ \gamma \leftrightarrow \delta \}  \right)\right], \nonumber\\
P^{1^-}_{3,\alpha\beta\gamma\delta}&=& \frac{1}{q^2}\,\left[
(g_{\alpha\gamma} q_\beta
q_\delta + \{ \alpha \leftrightarrow \beta \}) - 
\left(\{ \gamma \leftrightarrow \delta \}\right)\right],\nonumber\\
P^{1^+}_{4,\alpha\beta\gamma\delta}&=& 
\frac{1}{q^2}\,\left[
P^{1^-}_{4,\alpha\beta\gamma\delta} + q^2g_{\beta\gamma} g_{\alpha\delta}
  - q^2g_{\alpha\gamma} g_{\beta\delta}\right]\,.\nonumber 
\end{eqnarray*}
$P^{1^-}_3$ projects onto the twist-3 matrix element
$\kappa_3^\perp(K^*)$, $P^{1^-}_4$ onto $\kappa_4^\perp(K^*)$ and
$P^{1^+}_4$ onto the contribution from $1^+$ intermediate states. As
usual, $\Pi_{\alpha\beta\gamma\delta}$ must not have a singularity at $q^2=0$
which implies 
$$\Pi^{1^-}_{G,4}(0) + \Pi^{1^+}_{G,4}(0)=0.$$
That means that one can construct a mixed-parity sum rule from 
$\Pi_{G,4} \equiv 
(\Pi^{1^-}_{G,4}(q^2) + \Pi^{1^+}_{G,4}(q^2))/q^2$ which has lower
dimension than the pure-parity sum rule obtained from $\Pi^{1^-}_{G,4}$ alone.
We find\footnote{We also give $q^2$-independent terms in the quark
  condensate contribution to $\Pi_{G,4}^{1^\pm}$ because they are
  needed for calculating $\Pi_{G,4}$. Note that for
  $\Pi_{G,4}^{1^\pm}$ these terms are affected by finite counterterms as
  discussed in Ref.~\cite{BZ}, which however cancel in 
the sum $\Pi_{G,4}^{1^-}+\Pi_{G,4}^{1^+}$.}
\begin{eqnarray}
\lefteqn{\Pi^{1^-}_{G,4}(q^2) =  
(m_s^2-m_q^2)\,\frac{\alpha_s}{144\pi^3}\,q^2\left[3\ln^2\frac{-q^2}{\mu^2}-
11\ln\frac{-q^2}{\mu^2}\right]}\hspace*{1cm}\nonumber\\
&&{} + \frac{\alpha_s m_s}{3\pi}\left[
  \frac{5}{6}\,\squark+\left(\ln\,\frac{-q^2}{\mu^2} -
  \frac{5}{3}\right)\quark\right] - \frac{\alpha_s m_q}{3\pi}\left[
  \frac{5}{6}\,\quark+\left(\ln\,\frac{-q^2}{\mu^2} -
  \frac{5}{3}\right)\squark\right]
\nonumber \\
&&{}+ \frac{1}{12q^2}\,\mixed (2m_s+m_q)- \frac{1}{12
  q^2}\,\smixed(m_s+2 m_q)\nonumber\\
&&{}+\frac{1}{24 q^2}  \gluon\left\{ -m_s^2\left[ 2 -
\ln\frac{-q^2}{m_s^2} \right]+m_q^2 \left[ 2 -
\ln\frac{-q^2}{m_q^2} \right] + 2m_q m_s\ln\frac{m_q^2}{m_s^2}\right\}
\nonumber\\
&&{}+0\cdot (\langle \bar q q\rangle^2 - \langle \bar s s
\rangle^2),\label{pi1minus}
\end{eqnarray}
\begin{eqnarray}
\lefteqn{\Pi^{1^+}_{G,4}(q^2) = 
(m_s^2-m_q^2)\,\frac{\alpha_s}{144\pi^3}\,q^2\left[3\ln^2\frac{-q^2}{\mu^2}-
17\ln\frac{-q^2}{\mu^2}\right]}\hspace*{0.8cm}\nonumber\\
&&{} + \frac{\alpha_s m_s}{3\pi}\left[
  \frac{7}{6}\,\squark+\left(-\ln\,\frac{-q^2}{\mu^2} +
  \frac{4}{3}\right)\quark\right] - \frac{\alpha_s m_q}{3\pi}\left[
  \frac{7}{6}\,\quark+\left(-\ln\,\frac{-q^2}{\mu^2} +
  \frac{4}{3}\right)\squark\right]\nonumber\\
&&{}+ \frac{1}{12q^2}\,\mixed (2m_s-m_q)- \frac{1}{12
  q^2}\,\smixed(m_s-2 m_q)\nonumber\\
&&{}+\frac{1}{24 q^2}  \gluon\left\{ -m_s^2\left[ 2 -
\ln\frac{-q^2}{m_s^2} \right]+m_q^2 \left[ 2 -
\ln\frac{-q^2}{m_q^2} \right] - 2m_q m_s\ln\frac{m_q^2}{m_s^2}\right\}
\nonumber\\
&&{}+0\cdot (\langle \bar q q\rangle^2 - \langle \bar s s
\rangle^2),\label{pi1plus}
\end{eqnarray}
\begin{eqnarray}
\Pi_{G,4}(q^2) &=&
(m_s^2-m_q^2)\,\frac{\alpha_s}{72\pi^3} \left[3\ln^2\frac{-q^2}{\mu^2}-
14\ln\frac{-q^2}{\mu^2}\right]\hspace*{6.5cm}\nonumber\\
&&{} +
\frac{\alpha_s m_s}{9\pi q^2}\left[ 6 \squark-\quark\right] -
\frac{\alpha_s m_q}{9\pi q^2}\left[ 6 \quark-\squark\right]\nonumber\\
&&{}+\frac{1}{6 q^4}\left(m_q\mixed-m_s\smixed\right)\nonumber
\end{eqnarray}
\begin{eqnarray}
&&{}+\frac{1}{12 q^4}  \gluon \left\{ -m_s^2\left[ 2-
\ln\frac{-q^2}{m_s^2} \right]+m_q^2 \left[ 2 -
\ln\frac{-q^2}{m_q^2} \right]\right\}.
\label{pig4}
\end{eqnarray}

\section{Borel Transforms}

QCD sum rules are obtained from the Borel transforms of the
correlation functions listed in the previous section. Most of the
transforms are straightforward, except for those of expressions of
type $1/(q^2)^n \ln (-q^2/\mu^2)$, which can, however, be conveniently 
calculated using the formula
$$
\frac{1}{\pi}\,{\rm Im}(-q^2-i0)^\alpha =
\frac{s^\alpha}{\Gamma(-\alpha)\Gamma(1+\alpha)}\,\Theta(s)
$$
with $s=-q^2$. We then obtain, including continuum subtraction of
contributions from $s>s_0$,
\begin{eqnarray*}
{\cal B}^{\rm sub}_{M^2} \,\frac{1}{q^2}\,\ln\frac{-q^2}{\mu^2} &=&
\gamma_E -\ln\frac{M^2}{\mu^2} + \int_{s_0}^\infty
\frac{ds}{s}\,e^{-s/M^2},\\
{\cal B}^{\rm sub}_{M^2} \,\frac{1}{(q^2)^2}\,\ln\frac{-q^2}{\mu^2} &=&
\frac{1}{M^2}\left(1-\gamma_E +\ln\frac{M^2}{\mu^2} + M^2\int_{s_0}^\infty
\frac{ds}{s^2}\,e^{-s/M^2}\right).
\end{eqnarray*}


\begin{thebibliography}{99}

\bibitem{BBNS}
M. Beneke {\em et al.},
%``{QCD} factorization for B $\to$ pi pi decays: Strong phases and CP
Phys.\ Rev.\ Lett.\  {\bf 83} (1999) 1914
[arXiv:hep-ph/9905312].
%%CITATION = HEP-PH 9905312;%%

\bibitem{FFs}
E.~Bagan, P.~Ball and V.~M.~Braun,
%``Radiative corrections to the decay B $\to$ pi e nu and the heavy quark
%limit,''
Phys.\ Lett.\ B {\bf 417} (1998) 154
[arXiv:hep-ph/9709243];\\
%%CITATION = HEP-PH 9709243;%%
P.~Ball,
%``B $\to$ pi and B $\to$ K transitions from {QCD} sum rules on the light-cone
JHEP {\bf 9809} (1998) 005
[arXiv:hep-ph/9802394];\\
%%CITATION = HEP-PH 9802394;%%
P.~Ball and V.~M.~Braun,
%``Exclusive semileptonic and rare B meson decays in {QCD},''
Phys.\ Rev.\ D {\bf 58} (1998) 094016
[arXiv:hep-ph/9805422];\\
%%CITATION = HEP-PH 9805422;%%
 P.~Ball and R.~Zwicky,
  %``Improved analysis of B $\to$ pi e nu from QCD sum rules on the
  %light-cone,''
  JHEP {\bf 0110} (2001) 019
  [arXiv:hep-ph/0110115];
  %%CITATION = HEP-PH 0110115;%%
Phys.\ Rev.\ D {\bf 71} (2005) 014015
  [arXiv:hep-ph/0406232];
  %%CITATION = HEP-PH 0406232;%%
  %``B/(d,s) $\to$ rho, omega, K*, Phi decay form factors from light-cone sum
  %rules revisited,''
  Phys.\ Rev.\ D {\bf 71} (2005) 014029
  [arXiv:hep-ph/0412079];
  %%CITATION = HEP-PH 0412079;%%
%``$|$V(ub)$|$ and constraints on the leading-twist pion distribution amplitude
%from B $\to$ pi l nu,''
Phys.\ Lett.\ B {\bf 625}, 225 (2005)
[arXiv:hep-ph/0507076].
%%CITATION = HEP-PH 0507076;%%
\bibitem{elena} P.~Ball and M.~Boglione,
  %``SU(3) breaking in K and K* distribution amplitudes,''
  Phys.\ Rev.\ D {\bf 68}, 094006 (2003)
  [arXiv:hep-ph/0307337].
  %%CITATION = HEP-PH 0307337;%%

\bibitem{alex}   A.~Khodjamirian, T.~Mannel and M.~Melcher,
  %``Kaon distribution amplitude from QCD sum rules,''
  Phys.\ Rev.\ D {\bf 70} (2004) 094002
  [arXiv:hep-ph/0407226].
  %%CITATION = HEP-PH 0407226;%%

\bibitem{lenz}  V.~M.~Braun and A.~Lenz,
  %``On the SU(3) symmetry-breaking corrections to meson distribution
  %amplitudes,''
  Phys.\ Rev.\ D {\bf 70}, 074020 (2004)
  [arXiv:hep-ph/0407282].
  %%CITATION = HEP-PH 0407282;%%

\bibitem{BZ}
P.~Ball and R.~Zwicky,
%``SU(3) breaking of leading-twist K and K* distribution amplitudes: A
%reprise,''
Phys.\ Lett.\ {\bf B} {\em in press} [arXiv:hep-ph/0510338].
%%CITATION = HEP-PH 0510338;%%

\bibitem{lattbec}   D.~Becirevic {\it et al.}, 
  %``Coupling of the light vector meson to the vector and to the tensor
  %current,''
  JHEP {\bf 0305}, 007 (2003)
  [arXiv:hep-lat/0301020];\\
  %%CITATION = HEP-LAT 0301020;%%
V.~M.~Braun {\it et al.},
%``A lattice calculation of vector meson couplings to the vector and  tensor
%currents using chirally improved fermions,''
Phys.\ Rev.\ D {\bf 68} (2003) 054501
[arXiv:hep-lat/0306006].
%%CITATION = HEP-LAT 0306006;%%

\bibitem{Russians}
V.~L.~Chernyak, A.~R.~Zhitnitsky and I.~R.~Zhitnitsky,
%``Meson Wave Functions And SU(3) Symmetry Breaking,''
Nucl.\ Phys.\ B {\bf 204} (1982) 477
[Erratum-ibid.\ B {\bf 214} (1983) 547];
%%CITATION = NUPHA,B204,477;%%
%``Wave Functions Of The Mesons Containing S, C, B Quarks,''
Sov.\ J.\ Nucl.\ Phys.\  {\bf 38} (1983) 775
[Yad.\ Fiz.\  {\bf 38} (1983) 1277].
%%CITATION = SJNCA,38,775;%%

\bibitem{CZreport} 
V.~L.~Chernyak and A.~R.~Zhitnitsky,
%``Asymptotic Behavior Of Exclusive Processes In QCD,''
Phys.\ Rept.\  {\bf 112} (1984) 173.
%%CITATION = PRPLC,112,173;%%

\bibitem{BB98}
P.~Ball and V.~M.~Braun,
%``Higher twist distribution amplitudes of vector mesons in {QCD}: Twist-4
%distributions and meson mass corrections,''
Nucl.\ Phys.\ B {\bf 543} (1999) 201
[arXiv:hep-ph/9810475].
%%CITATION = HEP-PH 9810475;%%

\bibitem{PB98}
P.~Ball,
%``Theoretical update of pseudoscalar meson distribution amplitudes of  higher
%twist: The nonsinglet case,''
JHEP {\bf 9901}, 010 (1999)
[arXiv:hep-ph/9812375].
%%CITATION = HEP-PH 9812375;%%

\bibitem{BBKT}
P. Ball {\it et al.},
%``Higher twist distribution amplitudes of vector mesons in {QCD}: Formalism
Nucl.\ Phys.\ B {\bf 529} (1998) 323
[arXiv:hep-ph/9802299].
%%CITATION = HEP-PH 9802299;%%

\bibitem{prep} P. Ball, V.M.~Braun and A. Lenz, {\it in preparation}.

\bibitem{mslatt}
F.~Knechtli,
%``Lattice computation of the strange quark mass in QCD,''
arXiv:hep-ph/0511033.
%%CITATION = HEP-PH 0511033;%%

\bibitem{jamin}
E.~Gamiz {\it et al.},
%``V(us) and m(s) from hadronic tau decays,''
Phys.\ Rev.\ Lett.\  {\bf 94} (2005) 011803
[arXiv:hep-ph/0408044];\\
%%CITATION = HEP-PH 0408044;%%
S.~Narison,
%``Strange quark mass from e+ e- revisited and present status of light quark
%masses,''
arXiv:hep-ph/0510108.
%%CITATION = HEP-PH 0510108;%%

\bibitem{chPT}
H.~Leutwyler,
%``The ratios of the light quark masses,''
Phys.\ Lett.\ B {\bf 378} (1996) 313
[arXiv:hep-ph/9602366].
%%CITATION = HEP-PH 9602366;%%

\bibitem{mqlatt}
D.~Becirevic {\it et al.},
%``Non-perturbatively renormalised light quark masses from a lattice simulation
%with N(f) = 2,''
arXiv:hep-lat/0510014;\\
%%CITATION = HEP-LAT 0510014;%%
Q.~Mason {\it et al.} [HPQCD Collaboration],
%``High-precision determination of the light-quark masses from
%realistic lattice%QCD,''
arXiv:hep-ph/0511160.
%%CITATION = HEP-PH 0511160;%%

\bibitem{PDG}
S.~Eidelman {\it et al.}  [Particle Data Group Collaboration],
%``Review of particle physics,''
Phys.\ Lett.\ B {\bf 592} (2004) 1.
%%CITATION = PHLTA,B592,1;%%

\bibitem{logms}
S.~C.~Generalis and D.~J.~Broadhurst,
%``The Heavy Quark Expansion And QCD Sum Rules For Light Quarks,''
Phys.\ Lett.\ B {\bf 139} (1984) 85;\\
%%CITATION = PHLTA,B139,85;%%
V.~P.~Spiridonov and K.~G.~Chetyrkin,
%``Nonleading Mass Corrections And Renormalization Of The Operators M Psi-Bar
%Psi And G**2(Mu Nu),''
Sov.\ J.\ Nucl.\ Phys.\  {\bf 47} (1988) 522
[Yad.\ Fiz.\  {\bf 47} (1988) 818];\\
%%CITATION = SJNCA,47,522;%%
M.~Jamin and M.~M\"{u}nz,
%``Current correlators to all orders in the quark masses,''
Z.\ Phys.\ C {\bf 60} (1993) 569
[arXiv:hep-ph/9208201].
%%CITATION = HEP-PH 9208201;%%

\end{thebibliography}
\end{document}